\documentclass[twocolumn]{aastex62}

\usepackage{comment}

\usepackage{etoolbox}
\makeatletter
\patchcmd\linenumberpar{\@LN@parpgbrk}{\penalty\@LN@parpgpen\relax}{}{}
\makeatother

\usepackage{subcaption}
\usepackage{graphicx}	
\usepackage{amsmath}	
\usepackage{amssymb}	
\usepackage[para,flushleft]{threeparttable} 
\usepackage{epstopdf}
\usepackage{textgreek}
\usepackage{comment}
\usepackage{multirow}

\usepackage{xcolor, fontawesome}
\definecolor{twitterblue}{RGB}{64,153,255}
\newcommand{\twitter}[1]{\href{https://twitter.com/#1}{\textcolor{twitterblue}{\faTwitter}\,\tt\hspace{2pt}\textcolor{blue!60!black}{@#1}}}

\hypersetup{colorlinks,linkcolor={red!60!black},citecolor={blue!60!black},urlcolor={blue!60!black}}

\usepackage{natbib,twoopt}
\bibpunct{(}{)}{;}{a}{}{,} 
\makeatletter
\newcommandtwoopt{\citeads}[3][][]{\href{http://ui.adsabs.harvard.edu/abs/#3}%
{\def\hyper@linkstart##1##2{}%
\let\hyper@linkend\@empty\citealp[#1][#2]{#3}}}
\newcommandtwoopt{\citepads}[3][][]{\href{http://ui.adsabs.harvard.edu/abs/#3}%
{\def\hyper@linkstart##1##2{}%
\let\hyper@linkend\@empty\citep[#1][#2]{#3}}}
\newcommandtwoopt{\citetads}[3][][]{\href{http://ui.adsabs.harvard.edu/abs/#3}%
{\def\hyper@linkstart##1##2{}%
\let\hyper@linkend\@empty\citet[#1][#2]{#3}}}
\newcommandtwoopt{\citeyearads}[3][][]%
{\href{http://ui.adsabs.harvard.edu/abs/#3}
{\def\hyper@linkstart##1##2{}%
\let\hyper@linkend\@empty\citeyear[#1][#2]{#3}}}
\makeatother

\graphicspath{{./}{figures/}}

\accepted{February 14, 2022}
\submitjournal{ApJ}

\shorttitle{Space Weather in the AU Mic System: Stellar Winds and Extreme Coronal Mass Ejections} 
\shortauthors{Alvarado-G\'omez et al.}

\begin{document}

\title{Simulating the Space Weather in the AU Mic System: Stellar Winds and Extreme Coronal Mass Ejections}

\correspondingauthor{J. D. Alvarado-G\'omez}
\email{julian.alvarado-gomez@aip.de}

\author[0000-0001-5052-3473]{Juli\'an D. Alvarado-G\'omez}
\altaffiliation{Karl Schwarzschild Fellow $|$ \twitter{AstroRaikoh}}
\affil{Leibniz Institute for Astrophysics Potsdam, An der Sternwarte 16, 14482 Potsdam, Germany}

\author[0000-0003-3721-0215]{Ofer~Cohen}
\affil{University of Massachusetts at Lowell, Department of Physics \& Applied Physics, 600 Suffolk Street, Lowell, MA 01854, USA}

\author[0000-0002-0210-2276]{Jeremy~J.~Drake}
\affil{Center for Astrophysics $|$ Harvard \& Smithsonian, 60 Garden Street, Cambridge, MA 02138, USA}

\author[0000-0002-5456-4771]{Federico Fraschetti}
\affil{Center for Astrophysics $|$ Harvard \& Smithsonian, 60 Garden Street, Cambridge, MA 02138, USA}
\affil{Dept. of Planetary Sciences-Lunar and Planetary Laboratory, University of Arizona, Tucson, AZ, 85721, USA}

\author[0000-0003-1231-2194]{Katja Poppenhaeger}
\affil{Leibniz Institute for Astrophysics Potsdam, An der Sternwarte 16, 14482 Potsdam, Germany}
\affil{University of Potsdam, Institute for Physics and Astronomy, Karl-Liebknecht-Str. 24/25, 14476 Potsdam, Germany}

\author[0000-0002-8791-6286]{Cecilia Garraffo}
\affil{Center for Astrophysics $|$ Harvard \& Smithsonian, 60 Garden Street, Cambridge, MA 02138, USA}

\author[0000-0003-0695-6487]{Judy Chebly}
\affil{Leibniz Institute for Astrophysics Potsdam, An der Sternwarte 16, 14482 Potsdam, Germany}
\affil{University of Potsdam, Institute for Physics and Astronomy, Karl-Liebknecht-Str. 24/25, 14476 Potsdam, Germany}

\author[0000-0002-6299-7542]{Ekaterina Ilin}
\affil{Leibniz Institute for Astrophysics Potsdam, An der Sternwarte 16, 14482 Potsdam, Germany}
\affil{University of Potsdam, Institute for Physics and Astronomy, Karl-Liebknecht-Str. 24/25, 14476 Potsdam, Germany}

\author[0000-0001-7944-0292]{Laura Harbach}
\affil{Imperial College London, Astrophysics Group, Department of Physics, Prince Consort Rd, London, SW7 2AZ, UK}

\author[0000-0003-3061-4591]{Oleg Kochukhov}
\affil{Department of Physics and Astronomy, Uppsala University, Box 516, SE-75120 Uppsala, Sweden}



\begin{abstract}
\noindent Two close-in planets have been recently found around the M-dwarf flare star AU Microscopii (AU~Mic). These Neptune-sized planets (AU Mic b and c) seem to be located very close to the so-called ``evaporation valley" in the exoplanet population, making this system an important target for studying atmospheric loss on exoplanets. This process, while mainly driven by the high-energy stellar radiation, will be strongly mediated by the space environment surrounding the planets. Here we present an investigation on this last area, performing 3D numerical modeling of the quiescent stellar wind from AU~Mic, as well as time-dependent simulations describing the evolution of a highly energetic Coronal Mass Ejection (CME) event in this system. Observational constraints on the stellar magnetic field and properties of the eruption are incorporated in our models. We carry out qualitative and quantitative characterizations of the stellar wind, the emerging CMEs, as well as the expected steady and transient conditions along the orbit of both exoplanets. Our results predict an extreme space weather for AU~Mic and its planets. This includes sub-Alfv\'enic regions for the large majority of the exoplanet orbits, very high dynamic and magnetic pressure values in quiescence (varying within $10^{2} - 10^{5}$ times the dynamic pressure experienced by the Earth), and an even harsher environment during the passage of any escaping CME associated with the frequent flaring observed in AU~Mic. These space weather conditions alone pose an immense challenge for the survival of the exoplanetary atmospheres (if any) in this system.    
\end{abstract}



\section{Introduction} \label{sec:intro}

\noindent AU Microscopii (AU Mic) is the $10^{\rm th}$ brightest M-dwarf in the sky ($V_{\rm mag}~\simeq~8.81$, \citeads{2011AJ....142..138L}), and located at a distance of $\sim$\,$9.72$~pc from the Sun \citepads{2018yCat.1345....0G}. It is a young (22 Myr) M1V star with a mass of $M_{\bigstar} \simeq 0.50~M_\odot$ and a radius of $R_{\bigstar} \simeq 0.75~R_\odot$ (\citeads{2020Natur.582..497P} and references therein). The star is known to harbor a $\sim$\,$175$~au wide (from $35$~au to $210$~au in stellocentric distance, \citeads{2004Sci...303.1990K}, \citeads{2006ApJ...648..652S}) edge-on debris disk, characterized by clumpy outflows evolving on timescales of months to a few years (Boccaletti et al.~\citeyearads{2015Natur.526..230B}, \citeyearads{2018A&A...614A..52B}, \citeads{2020ApJ...889L..21G}). 

Reflecting its young age, AU Mic rotates relatively fast ($P_{\rm rot} \simeq 4.85$~d, \citeads{2012AcA....62...67K}), displaying one of the brightest X-ray coronae in the solar neighbourhood ($L_{\rm X}~=~5.5 \times 10^{29}$~erg~s$^{-1}$, \citeads{1999A&AS..135..319H}). Significant flare activity on AU Mic has been reported from multi-wavelength observations  (e.g.,~\citeads{2002ApJ...581..626R}, \citeads{2003AdSpR..32.1149M}, \citeads{2005A&A...431..679M}, \citeads{2020ApJ...891...80M}, \citeads{2021A&A...649A.177M}), including an energetic candidate flare-Coronal Mass Ejection (CME) event (\citeads{1994ApJ...435..449C}, \citeads{1999ApJ...510..986K}). Long-term photometric and spectroscopic monitoring provide evidence of a possible $\sim$\,$5$~yr chromospheric activity cycle in the star \citepads{2019MNRAS.483.1159I}. 

\citetads{2020Natur.582..497P} discovered a Neptune-sized planet (AU~Mic~b) in a $\sim$\,$8.5$~d period orbit, and proposed a second candidate planet (AU~Mic~c) further out in the system. With the aid of additional observations, \citetads{2021A&A...649A.177M} refined the fundamental parameters of AU~Mic~b ($R_{\rm b} = 1.05 \pm 0.04~R_{\rm Nep}$, $M_{\rm b} = 1.00 \pm 0.27~M_{\rm Nep}$, $a_{\rm b}~=~ 19.1 \pm 0.3~R_{\bigstar}$), and confirmed the presence of AU~Mic~c ($R_{\rm c} = 0.84 \pm 0.04~R_{\rm Nep}$, $ 0.13 < M_{\rm c}~[M_{\rm Nep}] < 1.46$) placing it in a $\sim$\,$18.86$~d period orbit ($a_{\rm c} = 29.0\pm3.0~R_{\bigstar}$). 

The planets of the AU Mic system appear to be located above the so-called ``radius gap'' in the exoplanet population (i.e.,~a~dearth of planets with radii close to $1.5$--$2.0~R_{\oplus} \simeq 0.4$--$0.5~R_{\rm Nep}$; \citeads{2017AJ....154..109F}). This feature, also known as the ``evaporation valley'', was predicted to be the result of atmospheric photo-evaporation due to high-energy radiation (EUV, X-ray) from the host star (Owen \& Wu~\citeyearads{2013ApJ...775..105O}, \citeyearads{2017ApJ...847...29O}). The primordial extended atmospheres of young exoplanets would make them appear larger compared to planets whose atmospheric layers would have already been removed, creating this bi-modal distribution in planet size. Close-in orbits, relatively low exoplanet masses, and elevated coronal emission, are expected to facilitate this process (see \citeads{2019AREPS..47...67O}). This makes AU~Mic a key object in the study of atmospheric loss, particularly as a potentially observable example of young exoplanets entering and moving across the evaporation valley, as well as an archetype for understating how environmental conditions determine the evolution of planetary systems. 

While important, high-energy photon radiation is not the only aspect influencing the evolution of exoplanetary atmospheres. Planets around late-type stars orbit within a magnetized and highly ionized plasma environment. These conditions are created by the magnetic field of the host star and are subject to stochastic and secular variations driven by the unceasing fluctuations of stellar magnetic activity (see \citeads{2019BAAS...51c.113D}). Known as ``space weather", the interplanetary plasma conditions have a palpable effect on planets, including magnetospheric compression due to stellar wind dynamic and magnetic pressures (e.g.,~\citeads{2015MNRAS.449.4117V}, \citeads{2016ApJ...833L...4G}, \citeads{2019ApJ...875L..12A}), the induction of currents and subsequent Joule heating of exospheres (e.g.,~Cohen~et~al.~\citeyearads{2018ApJ...856L..11C}, \citeyearads{2020ApJ...897..101C}, \citeads{2020ApJ...902L...9A}), and the erosion of planetary atmospheres via a range of different thermal and non-thermal processes (see~\citeads{2020JGRA..12527639G} for a recent review).

In this context, this study presents numerical simulations of the steady stellar wind from AU~Mic, as well as the expected environment generated by an energetic CME event. These results are used to evaluate the quiescent and transient space weather experimented by planets AU~Mic~b and AU~Mic~c. Our state-of-the-art models employ observed properties of the surface magnetic field of AU~Mic as boundary conditions, as well as eruption parameters compatible with the best CME candidate observed on this flare star. 

The outline of this article is as follows: models, boundary conditions, and numerical set-up are presented in Sect.~\ref{sec:Methods}. The stellar wind steady-state solutions, time-dependent CME simulations, and the expected space environment for the exoplanets of the AU~Mic system, are presented and discussed in Sect.~\ref{sec:results}. We conclude and summarise our findings in Sect.~\ref{sec:Conclusions}.

\section{Magnetically-Driven Numerical Models}\label{sec:Methods}

\noindent The simulations presented here are performed using the most recent version of the \href{http://csem.engin.umich.edu/tools/swmf/}{Space Weather Modeling Framework} (SWMF, \citeads{2018LRSP...15....4G}). This suite of physics-based models, originally developed for solar system science, allows a detailed numerical description of a wide range of spatial domains covering different objects of interest; from local simulations of convection to global models of planetary magnetospheres and the heliosphere (see T\'oth et al. \citeyearads{2005JGRA..11012226T}, \citeyearads{2012JCoPh.231..870T}). In recent years, models incorporated in the SWMF have been used by different authors to study the space environment of other late-type stars and their exoplanets (e.g., \citeads{2016ApJ...833L...4G}, \citeads{2016MNRAS.459.1907N}, \citeads{2017A&A...602A..39V}, \citeads{2020ApJ...897..101C}).

To investigate the expected space weather surrounding AU~Mic we consider two coupled models of the SWMF. These correspond to the Alfv\'en Wave Solar Model (AWSoM, \citeads{2014ApJ...782...81V}), and the \citetads{1999A&A...351..707T} flux-rope eruption model, both of which have been validated in multiple studies of the Sun-Earth environment (e.g., \citeads{2013ApJ...773...50J}, \citeads{2019ApJ...872L..18V}, \citeads{2019ApJ...887...83S}). 

\subsection{Corona and Stellar Wind Model}\label{sec:SC}

\noindent The dissipation of Alfv\'en waves in the upper chromosphere constitutes the main mechanism behind the coronal heating and the acceleration of the stellar wind within AWSoM. These contributions appear as additional source terms in the non-ideal magneto-hydrodynamic (MHD) momentum and energy relations, which are solved alongside the magnetic field induction and mass conservation equations. Electron heat conduction and radiative losses are also incorporated in the simulation.

As described by \citetads{2014ApJ...782...81V}, AWSoM-specific proportionality constants, together with the base plasma density ($n_{0} = 2\times10^{10}$~cm$^{-3}$) and temperature ($T_{0} = 5\times10^{4}$~K), are required as boundary conditions. For the models discussed here, we assume the same set of AWSoM parameters\footnote[1]{Corresponding to the Alfv\'en wave correlation length $L_{\perp}\sqrt{B} = 1.5\times10^{5}$~m~$\sqrt{\text{T}}$, and the scaling of the Alf\'ven wave Poynting flux $(S/B)_{\bigstar} = 1.1\times10^{6}$~W~m$^{-2}$~T$^{-1}$.} as in the M-dwarf simulations presented in \citetads{2020ApJ...895...47A}, adjusting the basic stellar properties to AU~Mic ($M_{\bigstar} = 0.5$~$M_{\odot}$, $R_{\bigstar} = 0.75$~$R_{\odot}$, $P_{\rm rot} = 4.85$~d; \citeads{2012AcA....62...67K}, \citeads{2020Natur.582..497P}).

In order to calculate the required propagation and reflection of Alfv\'en waves, AWSoM also uses the distribution of the radial magnetic field ($B_{\rm R}$) on the stellar surface as input. For simulations in the stellar regime, this information can be retrieved from spectropolarimetric observations and Zeeman-Doppler Imaging (ZDI) magnetic field reconstructions (\citeads{1997A&A...326.1135D}, \citeads{2002A&A...381..736P}). 

\begin{figure}[!t]
\centering 
\includegraphics[trim = 0.5cm 0.5cm 0.5cm 0.5cm, clip=true,width=0.24\textwidth]{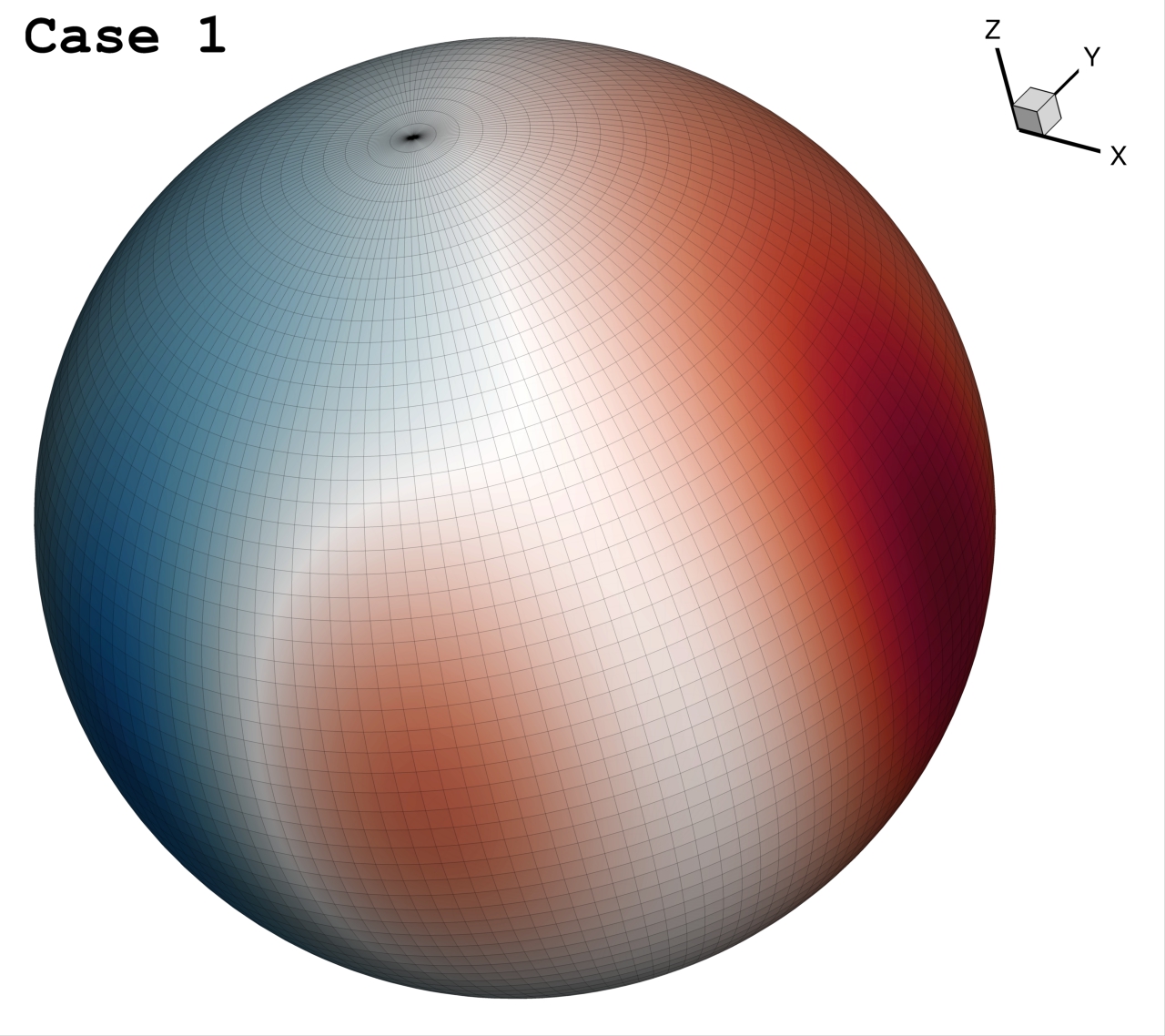}\includegraphics[trim = 0.5cm 0.5cm 0.5cm 0.5cm, clip=true,width=0.24\textwidth]{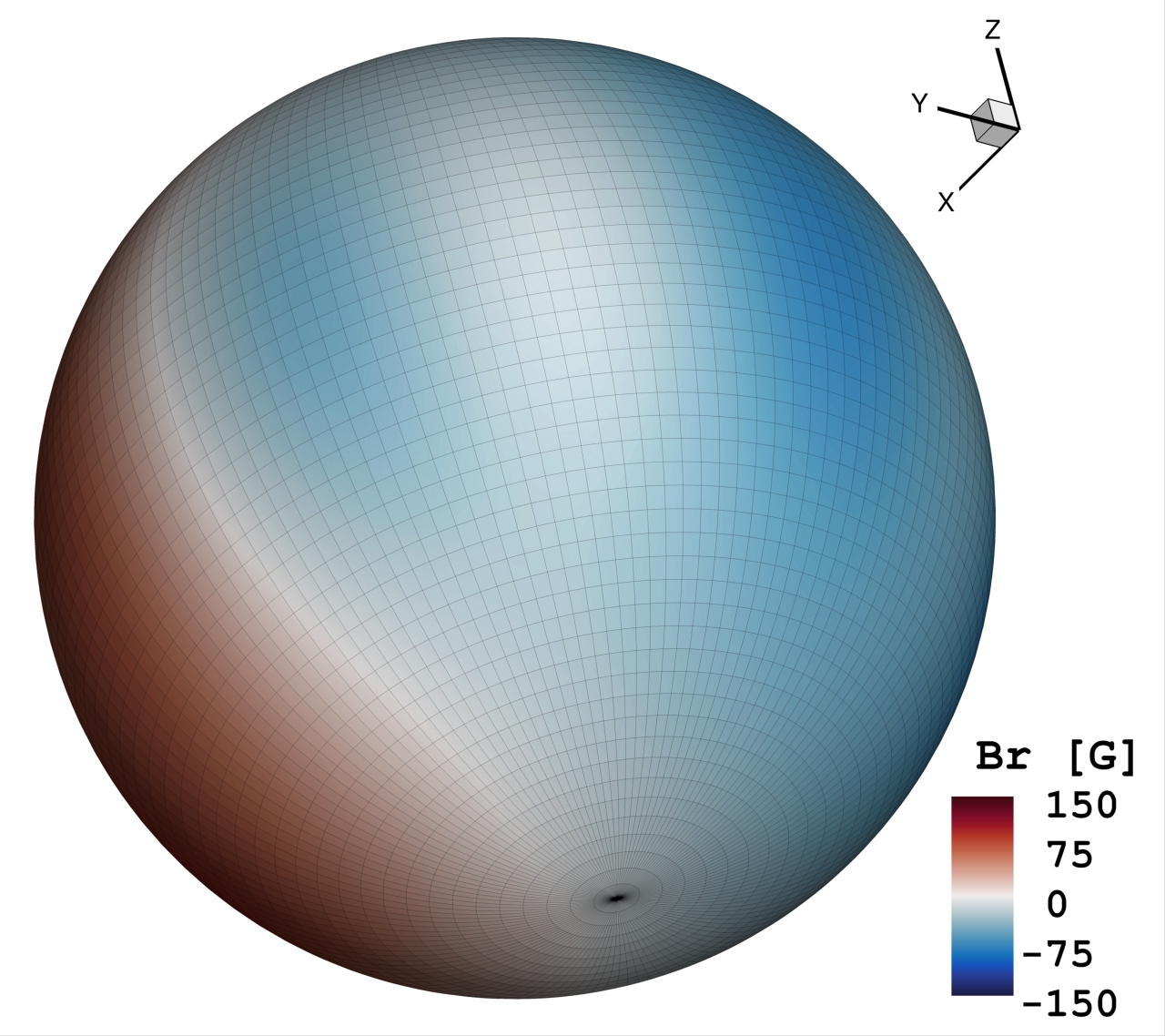}\vspace{0.45cm}
\includegraphics[trim = 0.5cm 0.5cm 0.5cm 0.5cm, clip=true,width=0.24\textwidth]{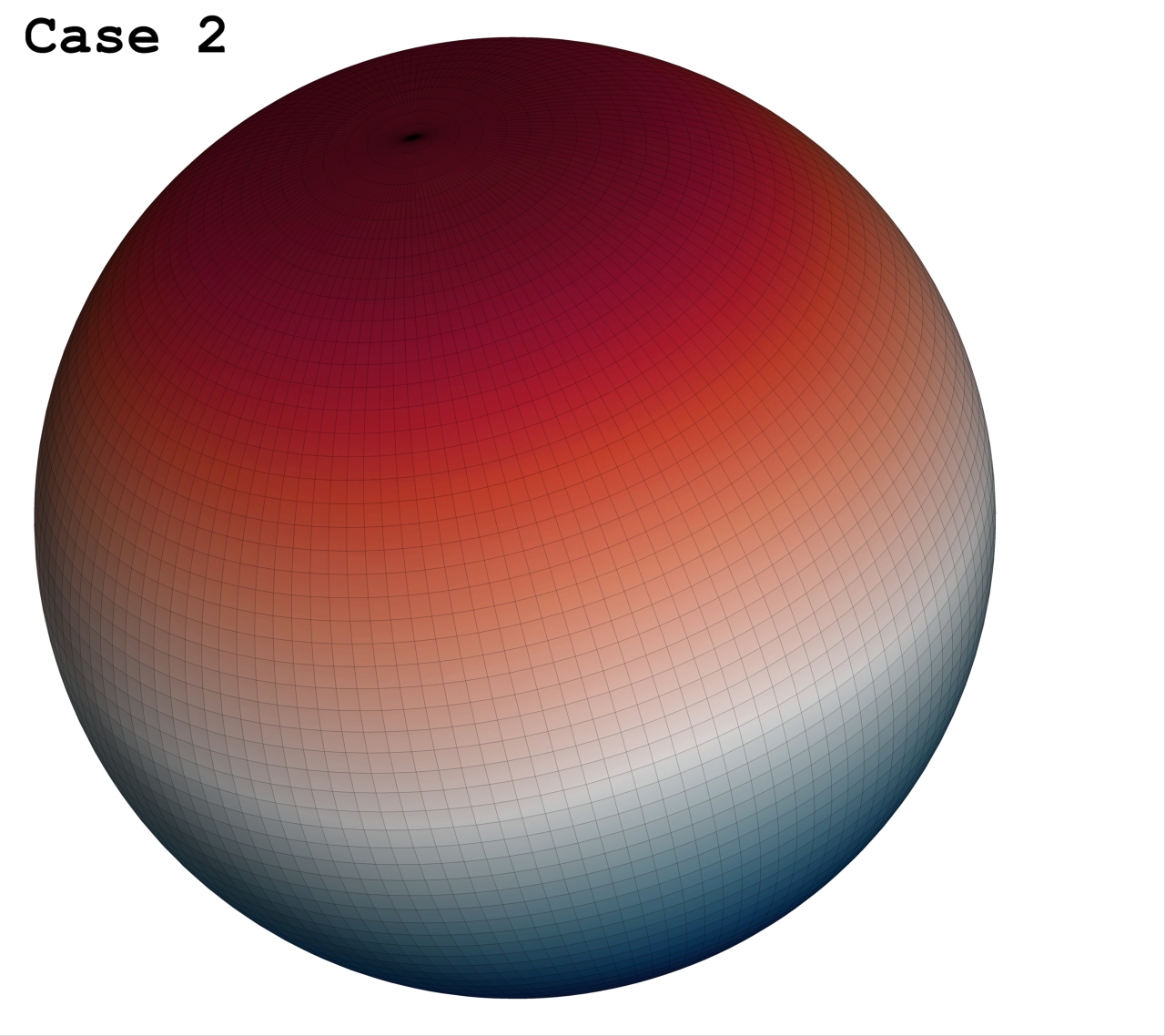}\includegraphics[trim = 0.5cm 0.5cm 0.5cm 0.5cm, clip=true,width=0.24\textwidth]{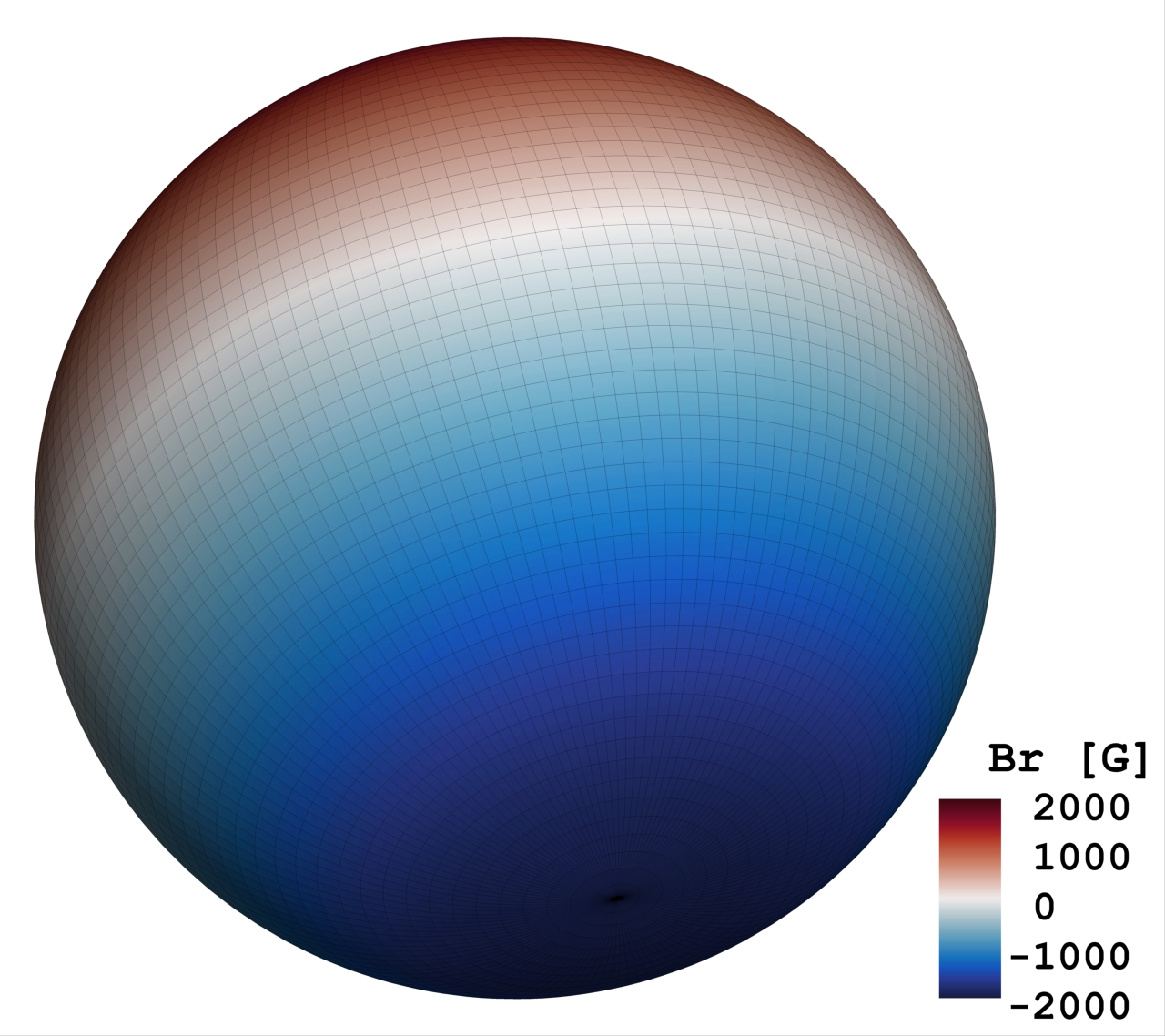}\vspace{0.45cm}
\includegraphics[trim = 0.5cm 0.5cm 0.5cm 0.5cm, clip=true,width=0.24\textwidth]{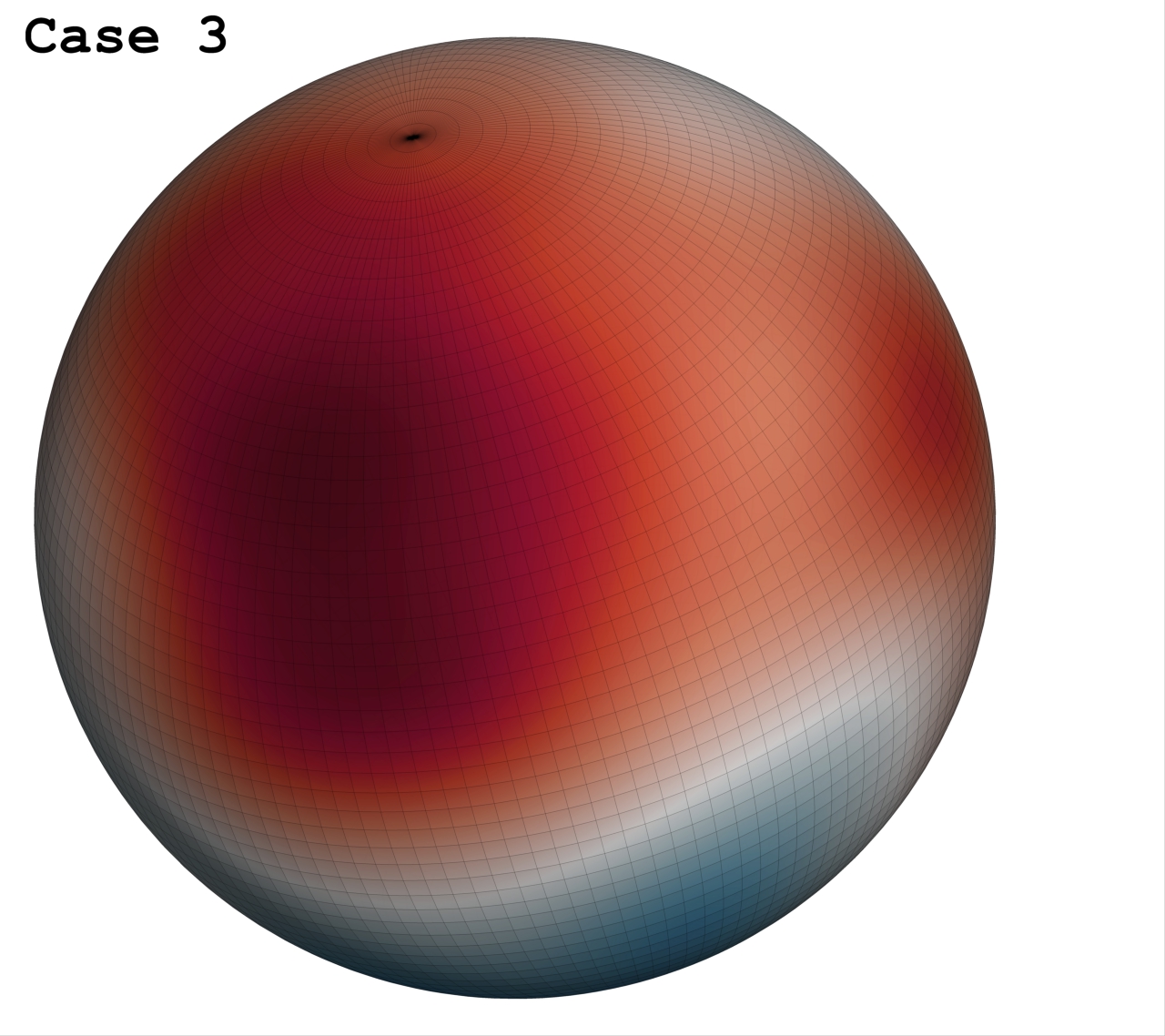}\includegraphics[trim = 0.5cm 0.5cm 0.5cm 0.5cm, clip=true,width=0.24\textwidth]{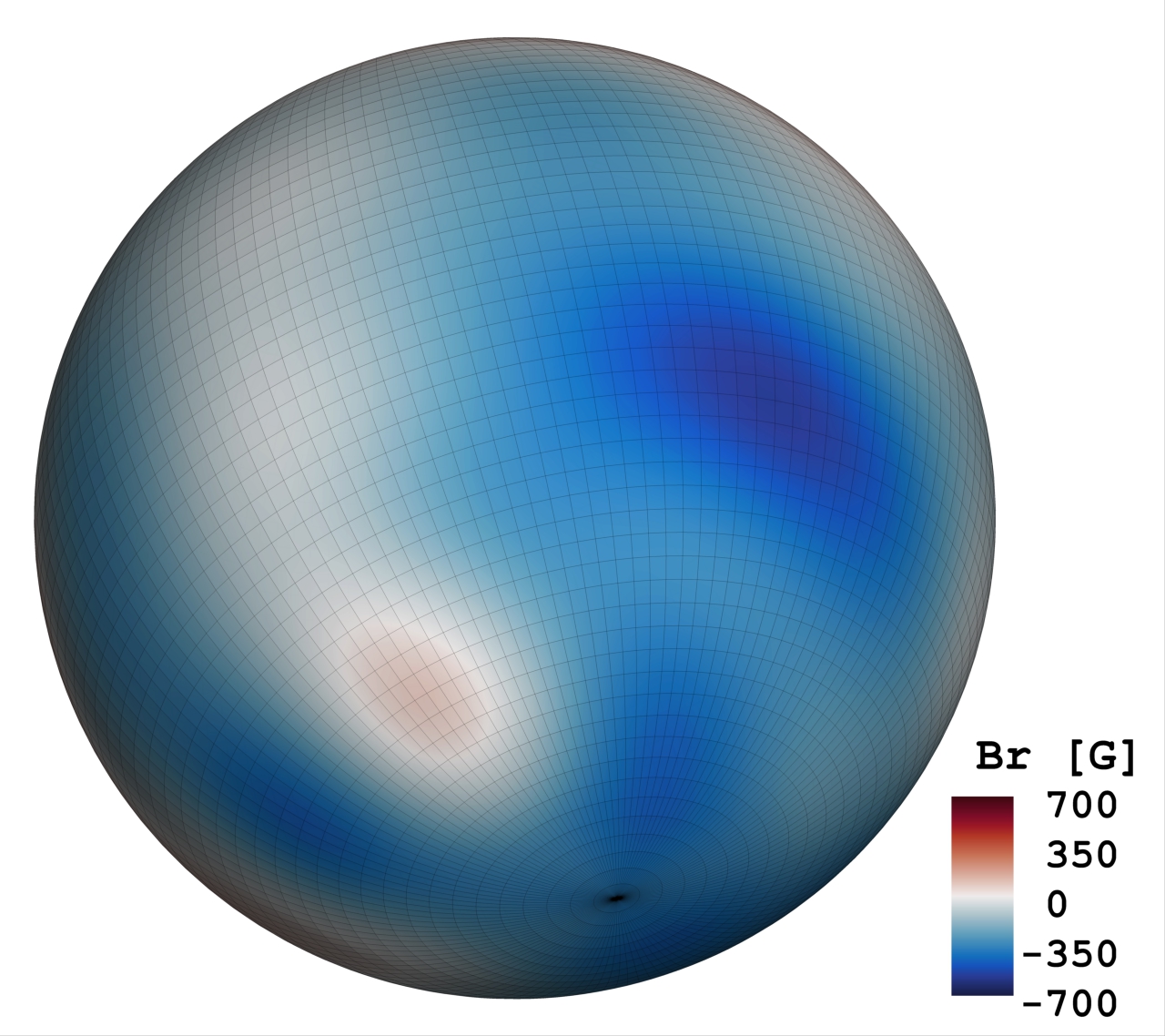}
\caption{North (left) and South (right) poleward views of the three radial magnetic field configurations driving our corona/stellar wind models for AU Mic. Case~1 is provided by the nominal ZDI reconstruction obtained \citetads{2020ApJ...902...43K}, while Case~2 includes an extra $2$~kG axisymmetric dipole (see text for details). Case~3 corresponds to the ZDI map reported by \citetads{2021MNRAS.500.1844K}. Note the different $B_{\rm R}$ range in each case.}\label{fig_ZDI}
\end{figure}

In the case of AU~Mic, two recent observational studies have provided information on the properties of its large-scale magnetic field. \citetads{2020ApJ...902...43K} obtained the first ZDI map of this star, revealing a non-axisymmetric field with a (signed) average strength of $\sim$\,$88$~G and values up to $\sim$\,$184$~G in local regions (Fig.~\ref{fig_ZDI}, Case~1). However, the authors caution that due to the particular line-of-sight orientation of AU Mic (close to $90^{\circ}$ of inclination), ZDI reconstructions employing circular polarization alone would not be sensitive to components anti-symmetric with respect to the stellar equator (e.g., a dipole aligned with the stellar rotation axis). To circumvent this limitation, \citetads{2020ApJ...902...43K} analyzed the Zeeman signatures from AU Mic in linear polarization, indicating the presence of an additional axisymmetric large-scale dipole field of $\sim$\,2.0~kG (Fig.~\ref{fig_ZDI}, Case~2). Employing a different set of circular polarization observations, \citetads{2021MNRAS.500.1844K} reported a second ZDI map of AU Mic. This map is characterized by a $450$~G large-scale dipole field inclined by $19^{\circ}$ with respect to the rotation axis, displaying an average longitudinal magnetic field strength of $475$~G (Fig.~\ref{fig_ZDI}, Case~3). The effects from the equator-on orientation of AU Mic were not discussed by \citetads{2021MNRAS.500.1844K} in their ZDI reconstruction. The radial magnetic field maps, incorporated at the inner boundary of our simulations, are presented in the different panels of Fig.~\ref{fig_ZDI}.

The grid in our simulations is spherical, with the inner boundary representing the stellar surface ($\sim$\,$1\,R_{\bigstar}$), and the rotation axis of the star aligned with the $z$ cartesian direction. The spatial resolution is higher at the surface ($\Delta R = 0.025\,R_{\bigstar}$, $\Delta \Phi = 1.4^{\circ}$\footnote[2]{Here $\Phi$ denotes both angular directions in spherical coordinates (azimuthal and meridional).}), with a radial stretching factor $\propto\ln(R)$. Given the varying maximum field strength of the input ZDI maps (see Fig.~\ref{fig_ZDI}), a different domain size is employed for each run: $80\,R_{\bigstar}$ (Case~1), $200\,R_{\bigstar}$ (Case~2), and $120\,R_{\bigstar}$ (Case~3). This is done to guarantee a closed Alfv\'en Surface (AS) of the stellar wind, necessary for a correct description of the escaping wind solution within each three-dimensional domain. From an initial finite difference potential field extrapolation \citepads{2011ApJ...732..102T}, the simulations evolve until a steady-state solution is reached.

\subsection{Flux-rope CME model}\label{sec_FR}

\noindent The SWMF implementation of the \citetads[TD,]{1999A&A...351..707T} flux-rope eruption model is used to perform our CME simulations. This model is based on a prescribed magnetic tension imbalance between the ambient magnetic field configuration (provided by the AWSoM steady-state solution) and a twisted magnetic loop inserted at the inner boundary of the domain (see~e.g.,~\citeads{2007SpWea...5.6003T}). The initialisation of the TD flux-rope requires 7 parameters describing its location (2), orientation, size (2), loaded mass ($M^{\rm FR}$) and magnetic free energy available in the eruption ($E_{\rm B}^{\rm FR}$).

Table~\ref{tab_1} contains the model parameters employed in our TD flux-rope CME simulation on AU~Mic. Both tilt angle and longitude were chosen in such a way that the inserted flux-rope would lie in the vicinity of a polarity inversion line on all three surface field configurations. For the anchoring latitude, we used one of the values of the starspot locations on AU~Mic derived from the photometric light curve modeling performed by \citetads{2019ApJ...883L...8W}. This selection was made so that the emerging CMEs would have a higher probability of impacting the planets lying in the equatorial plane of the system. The remaining parameters were adjusted so that the inserted TD flux-rope would be able to power a CME with mass and kinetic energy consistent with the best candidate event observed in this star so far ($M^{\rm CME} \sim 10^{20}$~g, $E_{\rm K}^{\rm CME} \sim 10^{36}$~erg; \citeads{1999ApJ...510..986K}).  

\begin{deluxetable}{lcc}[ht]
\tablecaption{Initialization parameters of the TD flux-rope CME simulation on AU~Mic.\label{tab_1}}
\tablecolumns{3}
\tablenum{1}
\tablehead{\colhead{Parameter} & \colhead{Value} & \colhead{Unit}}
\startdata
Tilt angle$^{\dagger}$ & 215.0 & deg\\
Longitude & 330.0 & deg\\
Latitude & 9.6 & deg\\
Radius ($R^{\rm FR}$) & 20.0 & Mm\\
Length ($L^{\rm FR}$) & 99.5 & Mm \\
Mass ($M^{\rm FR}$) & $10^{19}$ & g\\
Magnetic energy ($E_{\rm B}^{\rm FR}$) & $4.2 \times 10^{37}$ & erg\smallskip\\
\enddata
\tablenotetext{}{\hspace{-0.2cm}$^{\dagger }$Measured with respect to the stellar equator in the counter clock-wise direction.}
\end{deluxetable}

After the TD flux-rope insertion, the evolution of the resulting CME on each surface magnetic field configuration is followed by a time-dependent simulation covering 90 minutes (real-time), with full-domain snapshots extracted at a cadence of 1 minute. Note that while the spatial resolution in our simulations is not able to capture the internal structure of the flux-rope (6 resolution elements between the flux-rope ends at the surface), it is sufficient to describe its global evolution and the subsequent CME propagation within the domain. 

\section{Results and Discussion}\label{sec:results}

\noindent Below we present the results obtained for each of our ZDI-driven corona/stellar wind models of AU Mic. These are used to derive expected steady and transient conditions (due to an energetic CME event) around the star, and to evaluate the space environment experienced by the exoplanets of this system.

\subsection{Quiescent stellar wind conditions around AU~Mic}\label{sec:SS}

\begin{figure*}
\begin{subfigure}{0.495\linewidth}
\centering\includegraphics[trim = 0.2cm 0.2cm 0.2cm 0.2cm, clip=true,width=\textwidth]{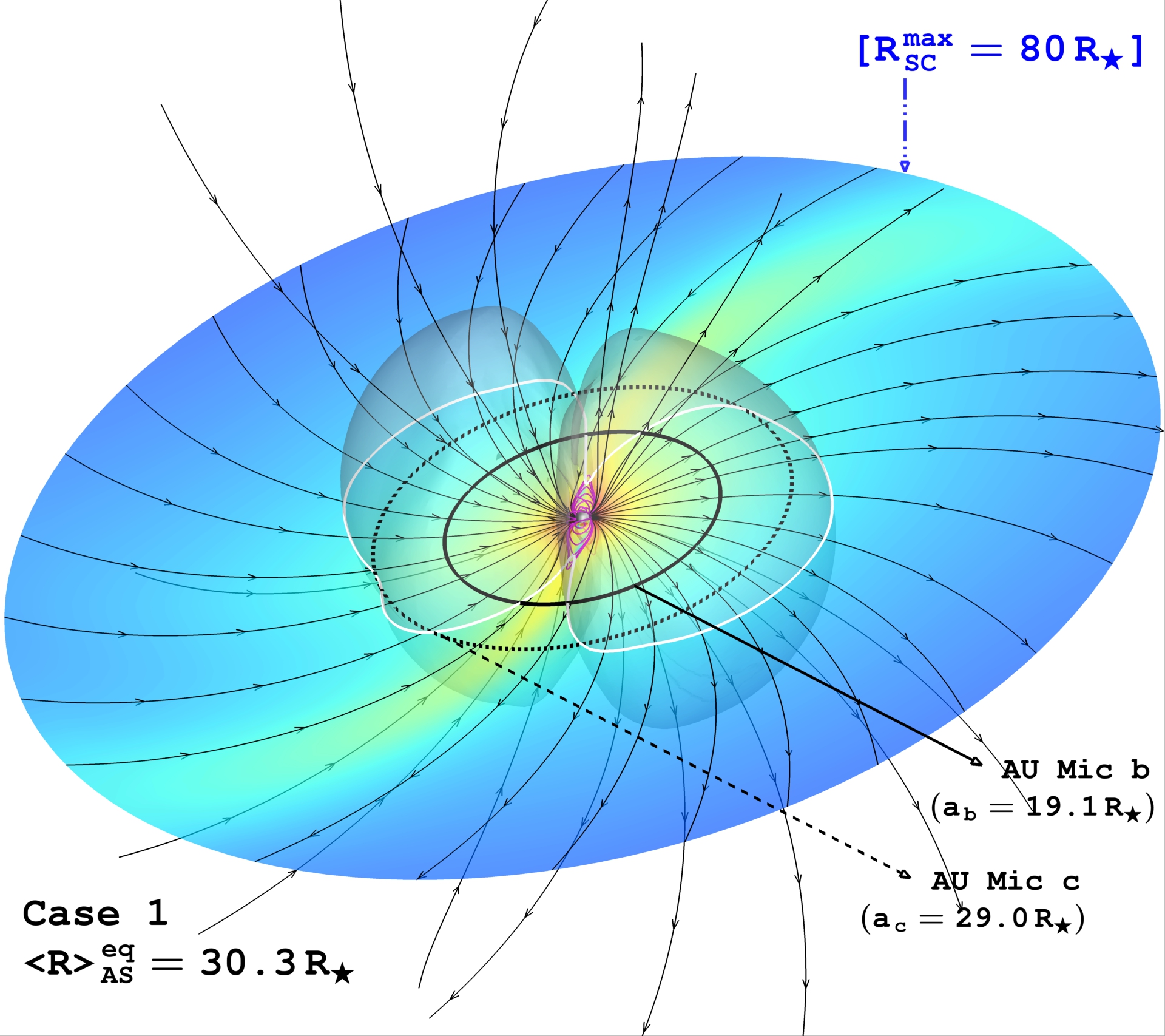}

\end{subfigure}
\begin{subfigure}{0.495\linewidth}
\centering\includegraphics[trim = 0.2cm 0.2cm 0.2cm 0.2cm, clip=true,width=\textwidth]{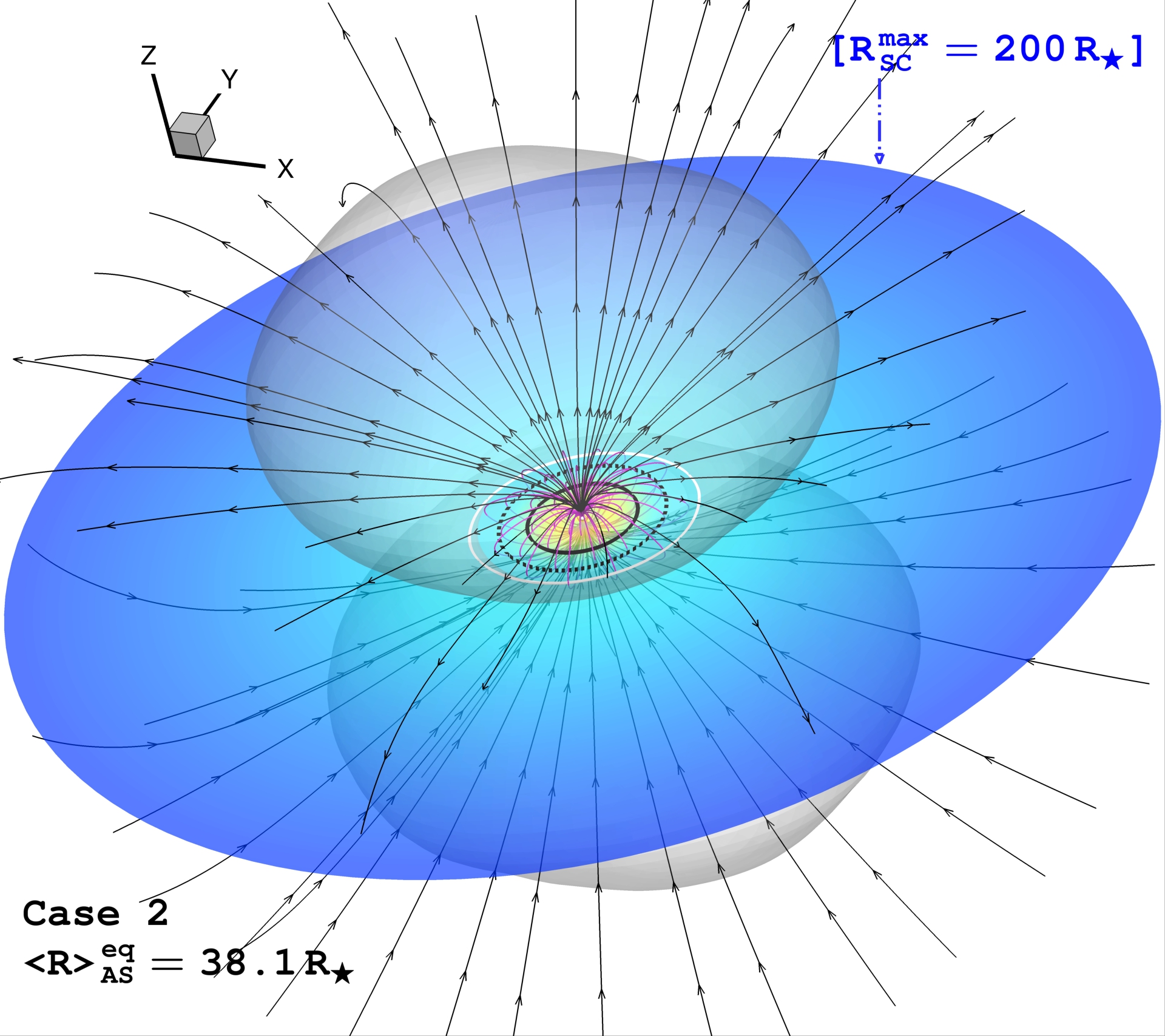}

\end{subfigure}\vspace{3pt}
 
\begin{subfigure}{0.495\linewidth}
\centering\includegraphics[trim = 0.2cm 0.2cm 0.2cm 0.2cm, clip=true,width=\textwidth]{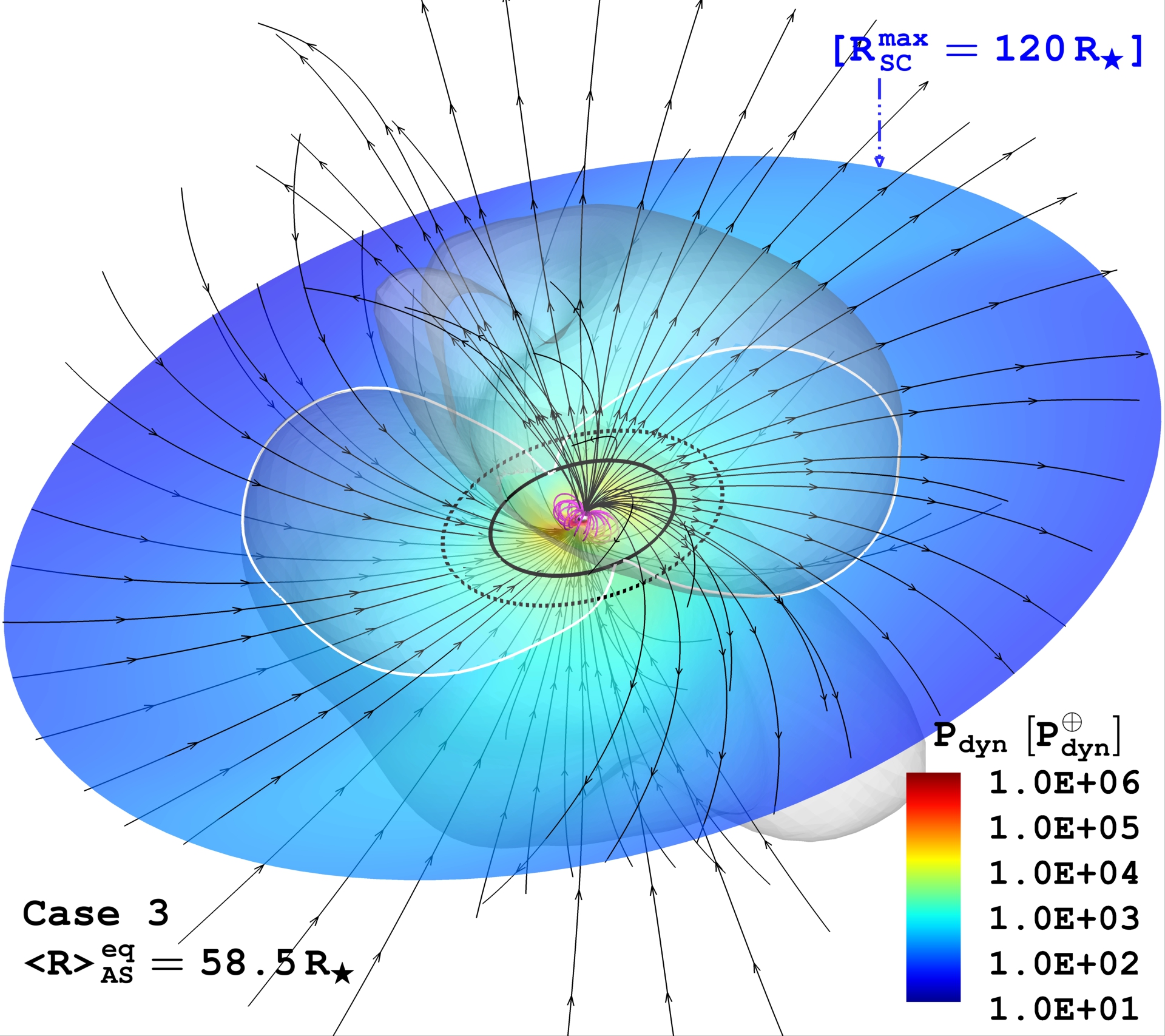}

\end{subfigure}%
\begin{subfigure}{0.495\linewidth}
\captionsetup{justification=centerlast,margin=0.25cm}
\captionof{figure}{\textbf{Figure 2:} Simulated stellar wind environment for AU~Mic driven by the different ZDI maps of the star (see Fig.~\ref{fig_ZDI}). A common color-scale denotes the equatorial distribution of the stellar wind dynamic pressure ($P_{\rm dyn}$) normalized to the average Sun-Earth value ($P^{\oplus}_{\rm dyn} = 1.5$~nPa). Solid and dashed lines indicate the orbits of AU~Mic~b and c, respectively. A translucent shade shows the resulting Alfv\'en Surface (AS) of the stellar wind in each case (see text for details). The equatorial projection of the AS (white solid line) and its average radius are indicated. Selected closed (magenta) and open (black with arrows) magnetic field lines are shown. The field of view on each panel has been adjusted to show the entire SC domain for the respective case.}
\end{subfigure}
\caption{}\label{Fig_SW}
\end{figure*}

\noindent The three panels of Fig.~\ref{Fig_SW} contain the steady-state stellar wind solutions obtained for AU~Mic, employing observed large-scale magnetic field distributions of this star as boundary conditions (Sect.~\ref{sec:Methods}). For each case we include the orbits of planets b and~c (Sect.~\ref{sec:intro}), as well as the AS of the stellar wind. This structure is defined by the spatial locations with an Alfv\'enic Mach number $M_{\rm A} = 1$, where the wind velocity matches the Alfv\'en speed $V_{\rm A}$ (i.e.~$M_{\rm A} \equiv U/V_{\rm A} = U\sqrt{4\pi\rho}/B = 1$, where $\rho$, $U$, and $B$ correspond to the local stellar wind density, velocity, and magnetic field strength, respectively). The AS establishes the boundary for the escaping stellar wind ($M_{\rm A} > 1$) and the magnetically-coupled outflows in the corona ($M_{\rm A} < 1$) that could still fall back to the star. The AS also separates the sub- and super-Alfv\'enic regimes of the stellar wind, which will influence dramatically the structure of any magnetosphere/ionosphere around an orbiting planet in the system (see \citeads{2014ApJ...790...57C} and references therein). 

In line with other stellar wind studies (e.g., \citeads{2014ApJ...783...55C}, \citeads{2014MNRAS.441.2361V}, \citeads{2015ApJ...798..116R}) stronger surface magnetic fields lead to faster stellar winds and larger AS structures. The terminal radial wind speeds\footnote[3]{Taken as the maximum value achieved at the outer boundary of the simulation.} within the respective simulation domains were roughly $1150$~km~s$^{-1}$ (at $80~R_{\bigstar}$, Case~1), $2200$~km~s$^{-1}$ (at $200~R_{\bigstar}$, Case~2), and $2000$~km~s$^{-1}$ (at $150~R_{\bigstar}$, Case~3). Similarly, we obtain average AS radii of $27.8~R_{\bigstar}$ (Case~1), $106.4~R_{\bigstar}$ (Case~2), and $73.1~R_{\bigstar}$ (Case~3). In terms of the stellar wind mass loss rates ($\dot{\rm M}_{\bigstar}$) our models yield values between $\sim$\,$5-10$~$\dot{\rm M}_{\odot}$\footnote[4]{Where $\dot{\rm M}_{\odot} \simeq 2 \times 10^{-14}$~M$_{\odot}$ yr$^{-1}$ $= 1.265 \times 10^{12}$~g~s$^{-1}$}.  

\captionsetup[figure]{labelformat=empty}

Unfortunately, there are no observational constraints on the stellar wind properties of AU Mic. Applying the observed coronal/disk properties in this system to the results of theoretical, numerical and semi-empirical studies from the literature (\citeads{2006A&A...455..987A}, \citeads{2006ApJ...648..652S}, \citeads{2013ApJ...772..149C}, \citeads{2015A&A...581A..97S}), leads to mass loss rate estimates $\dot{\rm M}_{\bigstar}\sim10 - 300~\dot{\rm M}_{\odot}$. Very recently, \citetads{2021MNRAS.504.1511K} simulated the stellar wind of AU~Mic using the ZDI map of \citetads{2021MNRAS.500.1844K} as a boundary condition (Case~3). They report solutions with low ($\dot{\rm M}_{\bigstar}=27~\dot{\rm M}_{\odot}$) and high ($\dot{\rm M}_{\bigstar}=590~\dot{\rm M}_{\odot}$) mass loss rates\footnote[5]{\citetads{2021MNRAS.504.1511K} assumed a different set of initial AWSoM parameters compared to the models presented here.}. It is good to note that our relatively low $\dot{\rm M}_{\bigstar}$ values appear close to estimates from stellar wind astrospheric detections for other M-dwarf stars \citepads{2021ApJ...915...37W}. Similarly, our resulting range of mass loss rate per unit surface area on AU~Mic ($\dot{\rm M}_{\bigstar}/A_{\bigstar} = \dot{\rm M}_{\bigstar}/4\pi R_{\bigstar}^2 \simeq 9 - 18~\dot{\rm M}_{\odot}/A_{\odot}$) is consistent\footnote[6]{Ly-$\alpha$ astrospheric constraints of $\dot{\rm M}_{\bigstar}$ are expected to be accurate within a factor of $2$ \citepads{2005ApJ...628L.143W}.} with the Ly-$\alpha$ astrospheric constraint available for EV~Lac (M3.5V, $P_{\rm rot} = 4.38$~d, $F_{\rm X} = 1.56\times10^{7}$~erg~cm$^{-2}$~s$^{-1}$, $\dot{\rm M}_{\bigstar}/A_{\bigstar} \simeq 9.8~\dot{\rm M}_{\odot}/A_{\odot}$; \citeads{2005ApJ...628L.143W}) which displays very similar rotation and coronal activity compared to AU~Mic ($P_{\rm rot} = 4.85$~d, $F_{\rm X} = 1.60\times10^{7}$~erg~cm$^{-2}$~s$^{-1}$).  

Both transiting planets in the AU Mic system lie very close to the equatorial plane (i.e.,~$i \simeq 90^{\circ}$, \citeads{2021A&A...649A.177M}). Fig.~\ref{Fig_SW} shows the projection of the AS (white solid line) as well as its average size on this plane $\left<R\right>^{\rm eq}_{AS}$. In all three cases, we find that for the majority (if not the totality) of their orbit, both planets remain inside the sub-Alfv\'enic regime of the stellar wind. As described by \citetads{2014ApJ...790...57C}, a planetary global magnetosphere within such sub-Alfv\'enic stellar wind will tend to be open (similar to the Alfv\'en wings configuration observed in the Galilean moons; see \citeads{1998JGR...10319843N}). Under those conditions, the magnetic field of the star and the planet could re-connect directly, impeding the generation of a bow shock structure (see~\citeads{2021arXiv210405968A}). This could modify drastically the amount of stellar wind energy dissipated into the planetary ionosphere/atmosphere \citepads{2018ApJ...856L..11C}, as well as the observable patterns of evaporative outflows from the exoplanets in the system \citepads{2021ApJ...913..130H}. 

Note also that the $2$~kG dipole aligned with the stellar rotation axis in Case~2 generates a \emph{smaller} $\left<R\right>^{\rm eq}_{AS}$ compared to the $450$~G inclined dipole from Case~3. This follows from a general result of magnetically-driven stellar winds models in which, for simple large-scale magnetic field geometries (i.e. dipole, quadrupole), the largest AS lobes are found right above the regions with the strongest magnetic field, while the smallest portions of the AS will emerge in the vicinity of the polarity inversion lines of the surface field (\citeads{2015MNRAS.449.4117V}, \citeads{2015ApJ...807L...6G}, \citeads{2016A&A...594A..95A}). In this way, a stellar dipole field aligned with the $z$-axis (as in the Case~2 simulation), will have a polarity inversion line coincident with the equatorial plane, so that the resulting stellar wind AS will be the smallest at latitude $0^{\circ}$. On the other hand, the $\sim$\,$20^{\circ}$ inclination of the large-scale dipole field in Case~3 is sufficient to make the large lobes of the AS cross the equatorial plane, making it much bigger at that particular latitude despite the weaker surface field. This example clearly illustrates that the magnetic field strength alone is insufficient to characterize the stellar wind environment of a planet-hosting star. It is also worth bearing in mind that secular change of the surface magnetic field due to the growth and decay of active regions, and perhaps also to a magnetic cycle, can change the location of the current sheet and the shape of the AS. 

The visualizations of Fig.~\ref{Fig_SW} also display the stellar wind dynamic pressure ($P_{\rm dyn}=\rho\,U^2$), normalized to typical conditions experienced by the Earth\footnote[7]{Corresponding to $\sim$\,$1.5~$nPa associated with a solar wind particle densities of $n \simeq 4-5$~cm$^{-3}$ and a velocity of $U \simeq 450$~km~s$^{-1}$.} ($P^{\oplus}_{\rm dyn}$). Within the three considered cases, we find values of $P_{\rm dyn}$ along the orbit of AU Mic~b ranging between $\sim$\,$1150 - 13300~P^{\oplus}_{\rm dyn}$, going down to $\sim$\,$450 - 6800~P^{\oplus}_{\rm dyn}$ for AU~Mic~c. Similar conditions have been predicted for the habitable zone planet Proxima~b \citepads{2016ApJ...833L...4G}, as well as for the exoplanet candidate Proxima~d \citepads{2020ApJ...902L...9A}. Still, we find that the planetary environment around AU~Mic is much harsher than for Proxima Centauri. The reason for this is that the relatively strong large-scale surface magnetism of AU~Mic leads to magnetic pressure values ($P_{\rm mag} = B^{2}/8\pi \propto 1/R^{6}$) greater than the associated $P_{\rm dyn}$ at the exoplanet orbits (which is not the case for Proxima~b). At the distance of AU~Mic~b, we find that, on average, $P_{\rm mag}$ appears larger than $P_{\rm dyn}$ by factors of $1.4$ (Case~1), $11.1$ (Case~2), and $5.2$ (Case~3). Likewise, along the orbit of AU~Mic~c, Cases 2 and 3 show average $P_{\rm mag}$ values greater than $P_{\rm dyn}$ by factors of $2.3$ and $2.0$, respectively. Only in the simulation with the weakest surface magnetic field (Case~1), the average $P_{\rm dyn}$ dominates over $P_{\rm mag}$ at the distance of AU~Mic~c (by a factor of $\sim$\,$2.7$). The quiescent environment obtained for the AU~Mic planets appears then more comparable to the extreme conditions expected for the planets of the TRAPPIST-1 system \citepads{2017ApJ...843L..33G}.

\subsection{Simulated CME events in the AU Mic System}\label{sec:CMEs}

\noindent The steady-state stellar wind solutions discussed in the previous section serve as the initial state for our time-dependent CME simulations. The latter are driven by a single erupting TD flux-rope, initialized with the parameters provided in Sect.~\ref{sec_FR}. As expected, while the properties of the inserted flux-rope are identical, the varying magnetic/stellar wind conditions among the three cases lead to quite different CMEs. 

We follow the procedure introduced by Alvarado-G\'omez~et~al~(\citeyearads{2019ApJ...884L..13A}, \citeyearads{2020ApJ...895...47A}) to identify and characterise our simulated CME events. For each solution, we compute the 3D density contrast $n(t)/n^{\rm SS}$, with $n(t)$ and $n^{\rm SS}$ as the instantaneous and pre-CME (steady-state) particle density distributions, respectively. The escaping CME front is traced by the collection of points on the time-evolving isosurface $n(t)/n^{\rm SS} = 3.0$ with radial speeds greater than the local escape velocity\footnote[8]{Given by $v_{\rm esc} = \sqrt{2GM_{\bigstar}/H}$, where $H$ is the front height above the stellar surface and $G$ is the gravitational constant.}. As different sections of the CME face dissimilar coronal and stellar wind conditions, the front expansion is highly inhomogeneous. To estimate the instantaneous maximum radial speed of the CME as a whole ($U^{\rm CME}_{\rm R, T}$), we take the mean position among the $1\%$ outermost grid points forming the escaping front and determine the relative variation of this average value between consecutive frames ($1$~min cadence). The escaping CME front also serves as the limit of integration for a volume integral of the plasma density, allowing the determination of the total mass ($M^{\rm CME}_{\rm T}$) and kinetic energy ($E_{\rm K, T}^{\rm CME}$) of the eruption at each time-step. 

\captionsetup[figure]{labelformat=empty}
\begin{figure}[!t]
\center 
\includegraphics[trim = 0.1cm 0.1cm 0.1cm 0.1cm, clip=true,width=0.475\textwidth]{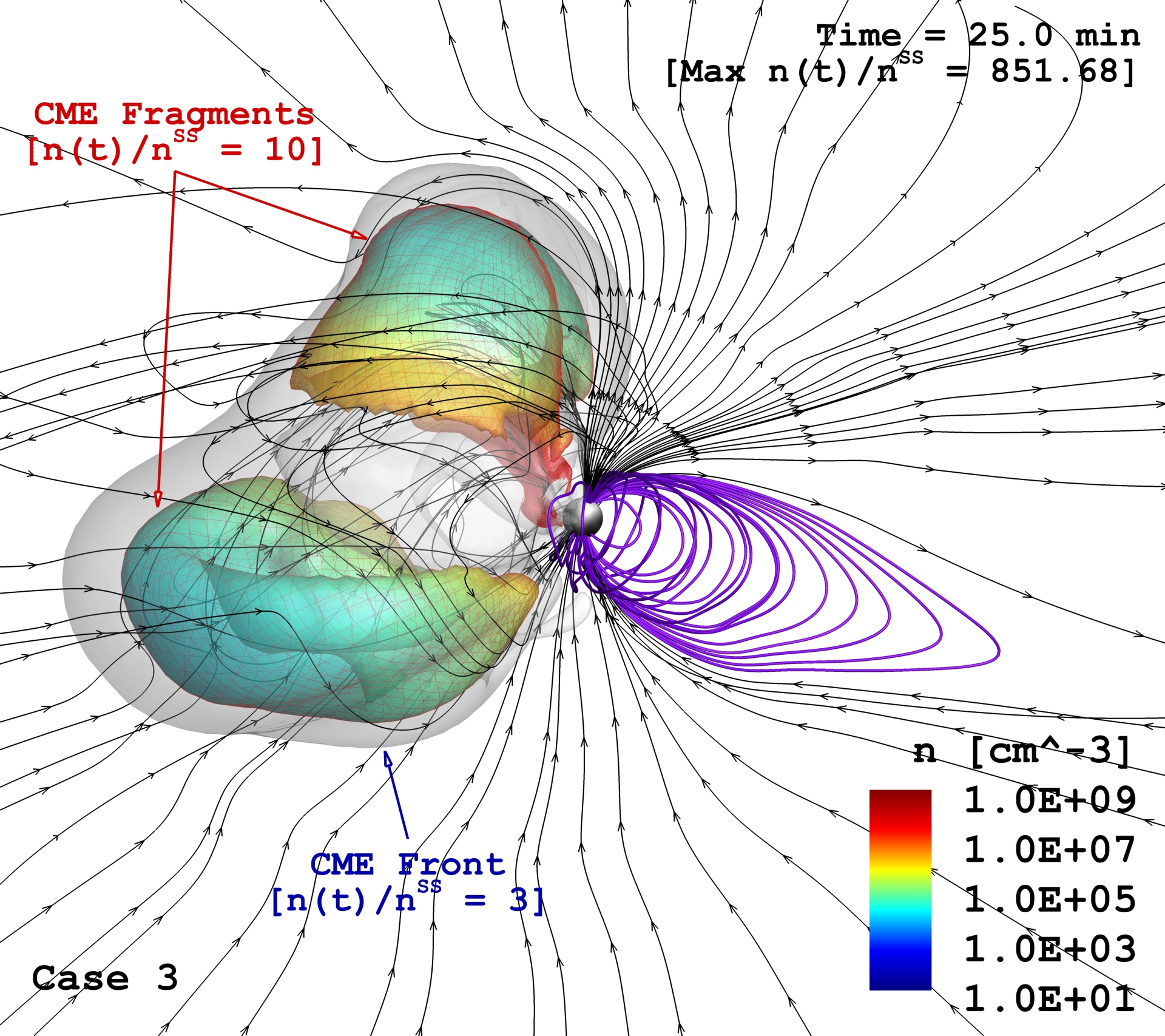}
\caption{}\label{Fig_Fragments}\vspace{-0.6cm}
\captionof{figure}{\textbf{Figure 3:} Example of the CME fragmentation that occurs in the Case~3 simulation. The instantaneous to steady-state density ratio, $n(t)/n^{\rm SS} = 3$, is used to identify the escaping CME structure (translucent isosurface) which envelops two spatially-disconnected regions where the contrast in density is larger than one order of magnitude. The boundary of these CME fragments is taken as the $n(t)/n^{\rm SS} = 10$ isosurface, which is color-coded by the local plasma density ($n$) to emphasise their different properties and dynamics. Selected magnetic field lines in the far (purple) and  erupting (black with arrows) sides are shown. The field of view of the visualization is $60~R_{\bigstar}$.}
\end{figure}

This analysis revealed that during their evolution, the emerging CME structures enclosed two separated regions where the density contrast increased by more than one order of magnitude (i.e., $n(t)/n^{\rm SS} > 10$). Figure~\ref{Fig_Fragments} shows the CME fragmentation in Case~3, but it was also observed in the other runs as well. As can be seen in the visualization, despite having a launching latitude of only $\sim$\,$10^{\circ}$ (see Table~\ref{tab_1}), one of the CME fragments follows a high-latitude trajectory (close to the north pole of the star) whereas the second one propagates closer to the equator. A similar behaviour was obtained on the strongest CME events analysed in \citetads{2018ApJ...862...93A}, where the stellar large-scale field only provided a weak magnetic confinement. The Case~2 simulation, in which the large-scale dipole field is the strongest of all cases (see Fig.~\ref{fig_ZDI}), shows that the two CME fragments are slowly guided in opposite poleward directions. Interestingly, only in the simulation for Case~1, where the surface field lacks a large-scale dipole field (see~Sect.~\ref{sec:SC}), is the propagation of both CME fragments centered along the equatorial plane where the planets reside. These results are better illustrated Figs.~\ref{Fig_CME1} and~\ref{Fig_CME2}, where the different snapshots correspond to the arrival time of the event at the orbit of AU~Mic~b (Fig.~\ref{Fig_CME1}) and AU~Mic~c (Fig.~\ref{Fig_CME2}).

\begin{deluxetable*}{CC|CCC|CCCCCC}[ht!]
\tablecaption{Properties obtained in the simulated CME events on AU~Mic. Parameters for the events as a whole ($\rm T$) as well as for the individual CME fragments ($1,2$) are included (see text for details). Values listed in columns $3-11$ correspond to temporal averages of instantaneous maxima during the entire event evolution.\label{tab_2}}
\tablecolumns{10}
\tablenum{2}
\tablehead{
\colhead{Case} & \colhead{\# of CME} & \colhead{$\left<M_{\rm T}^{\rm CME}\right>$} & \colhead{$\left<U_{\rm R,\,T}^{\rm CME}\right>$} & \colhead{$\left<E_{\rm K,\,T}^{\rm CME}\right>$} & \colhead{$\left<M_{\rm 1}^{\rm CME}\right>$} & \colhead{$\left<U_{\rm R,\,1}^{\rm CME}\right>$} & \colhead{$\left<E_{\rm K,\,1}^{\rm CME}\right>$} & \colhead{$\left<M_{\rm 2}^{\rm CME}\right>$} & \colhead{$\left<U_{\rm R,\,2}^{\rm CME}\right>$} & \colhead{$\left<E_{\rm K,\,2}^{\rm CME}\right>$}\vspace{-0.15cm}\\
\colhead{} & \colhead{fragments$^{\dagger}$} & \colhead{[$10^{18}$~g]} & \colhead{[km~s$^{-1}$]} & \colhead{[$10^{35}$~erg]} & \colhead{[$10^{18}$~g]} & \colhead{[km~s$^{-1}$]} & \colhead{[$10^{35}$~erg]} & \colhead{[$10^{18}$~g]} & \colhead{[km~s$^{-1}$]} & \colhead{[$10^{35}$~erg]}  
}
\startdata
1 & 2 & 2.15 & 10146 & 9.45 & 1.46 & 9924 & 6.24 & 0.56 & 4112 & 0.41\\
2 & 2 & 1.94 & 5102 & 3.51 & 0.95 & 4078 & 0.73 & 0.27 & 4968 & 0.65\\
3 & 2 & 2.51 & 9653 & 12.8 & 2.19 & 8278 & 8.53 & 0.09 & 9631 & 0.39\\
\enddata
\tablenotetext{}{$^{\dagger }$Identified by spatially-disconnected escaping perturbations with $n(t)/n^{\rm SS} \geq 10$.}
\end{deluxetable*}

To quantify parameters such as the speed ($U_{\rm R,\,1-2}^{\rm CME}$), mass ($M_{\rm 1-2}^{\rm CME}$) and kinetic energy ($E_{\rm K,\,1-2}^{\rm CME}$) of these CME fragments, we followed the methodology applied to the event as a whole, taking in this case a $n(t)/n^{\rm SS} = 10$ isosurface as the boundary of reference. Table~\ref{tab_2} contains time-averages of the CME properties for the entire event, as well as values associated with the individual CME fragments on each run. Notice that in all cases, the global values of $\left<M_{\rm T}^{\rm CME}\right>$ are close to two orders of magnitude lower than the CME candidate event parameters reported on AU Mic by \citetads{1999ApJ...510..986K}, and roughly five times lower than the amount of mass originally inserted in the TD flux-rope (see Table~\ref{tab_1}). This indicates that an important fraction of the flux-rope eruption does not escape and remains magnetically-confined in the low corona (see \citeads{2019ApJ...884L..13A}). Despite this, the escaping eruptions display $\left<E_{\rm K,\,T}^{\rm CME}\right>$ values consistent with the observed CME candidate event in this star.

In further agreement with the CME magnetic suppression results of \citetads{2018ApJ...862...93A}, the mean total mass of the eruptions are comparable between Cases 1 and 2, whilst the presence of a strong surface large-scale magnetic field in the latter case leads to an important reduction in the CME radial speed ($\sim$\,50\%) and kinetic energy ($\sim$\,65\%). Interestingly, the Case~3 simulation (driven by a different geometry which includes an inclined $450$~G large-scale dipole field; see Sect.~\ref{sec:SC}) only shows a minor reduction in $\left<U_{\rm R,\,T}^{\rm CME}\right>$ compared to Case~1 (by $\sim$\,5\%). While further investigation is required, this suggests that the large-scale dipole field inclination will strongly influence the dynamical properties of an emerging CME. 

In terms of the CME fragmentation, our calculations indicate that at least one fragment propagates with a similar speed to the global CME, and that combined they carry a significant fraction of the total escaping CME mass (between $60\%$ and $90\%$). However, the mass distribution among them is not uniform, with one fragment being considerably more massive than the other (see Table~\ref{tab_2}). For Cases 1 and 3, this mass discrepancy translates to a difference of more than one order of magnitude in the average kinetic energy of each CME fragment. The situation is more balanced in Case~2, as the lighter CME fragment compensates by traveling at a higher speed.  

As can be seen from Figs.~\ref{Fig_CME1} and \ref{Fig_CME2}, the impact of the CME on the exoplanets in the system will be mediated by the fragmentation of the eruption. Depending on the large-scale magnetic field geometry, both exoplanets can be strongly affected by one or two CME fragments (Cases 1 and 3), or can receive a relatively weak perturbation compared with the global energetics of the entire event (Case~2). In the following section, we evaluate how the space weather environment at the exoplanet orbits is affected by the arrival and passing of these CMEs.

\captionsetup[figure]{labelformat=empty}
\begin{figure*}
\begin{subfigure}{0.495\linewidth}
\centering\includegraphics[trim = 0.1cm 0.1cm 0.1cm 0.1cm, clip=true,width=\textwidth]{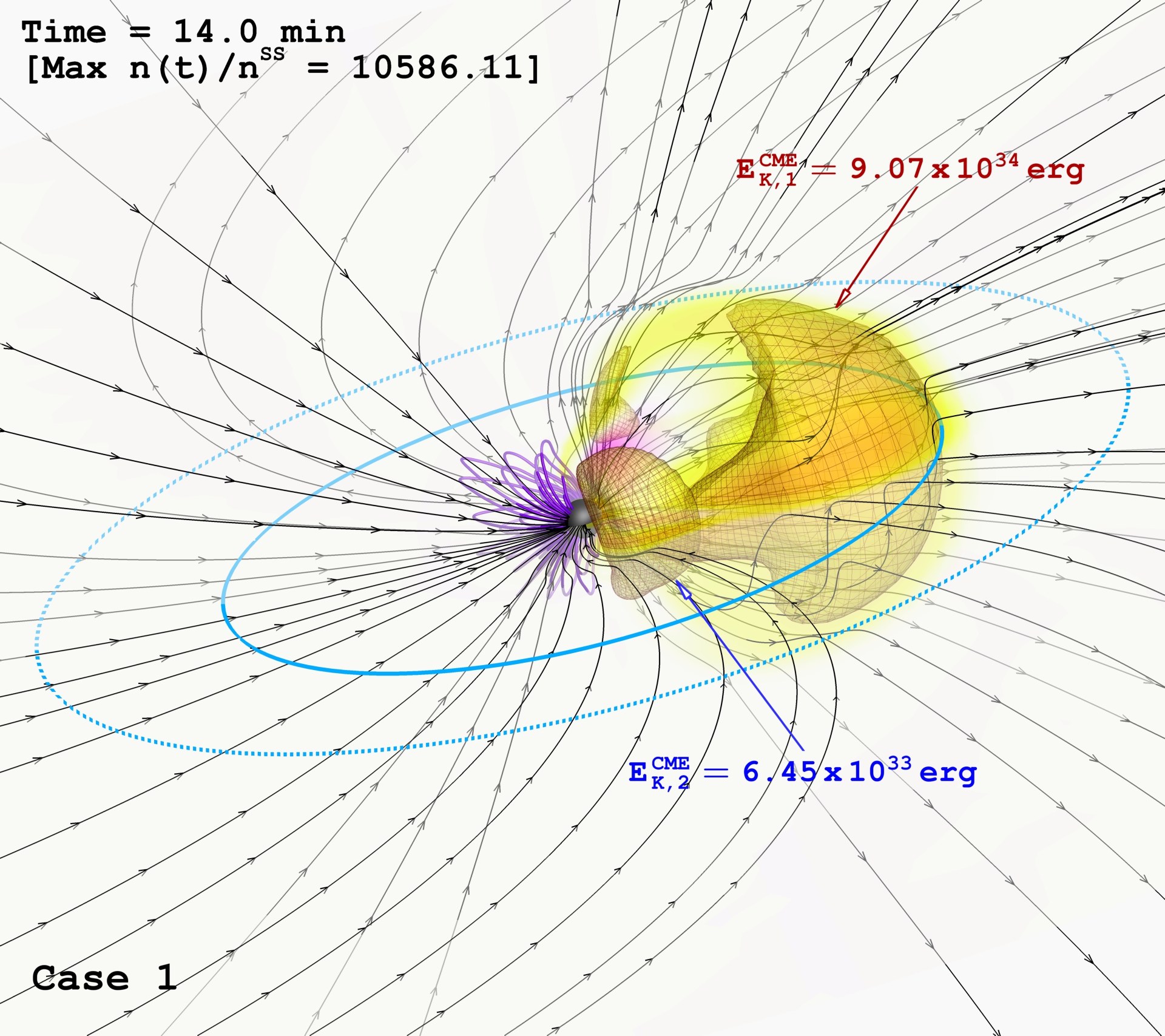}

\end{subfigure}
\begin{subfigure}{0.495\linewidth}
\centering\includegraphics[trim = 0.1cm 0.1cm 0.1cm 0.1cm, clip=true,width=\textwidth]{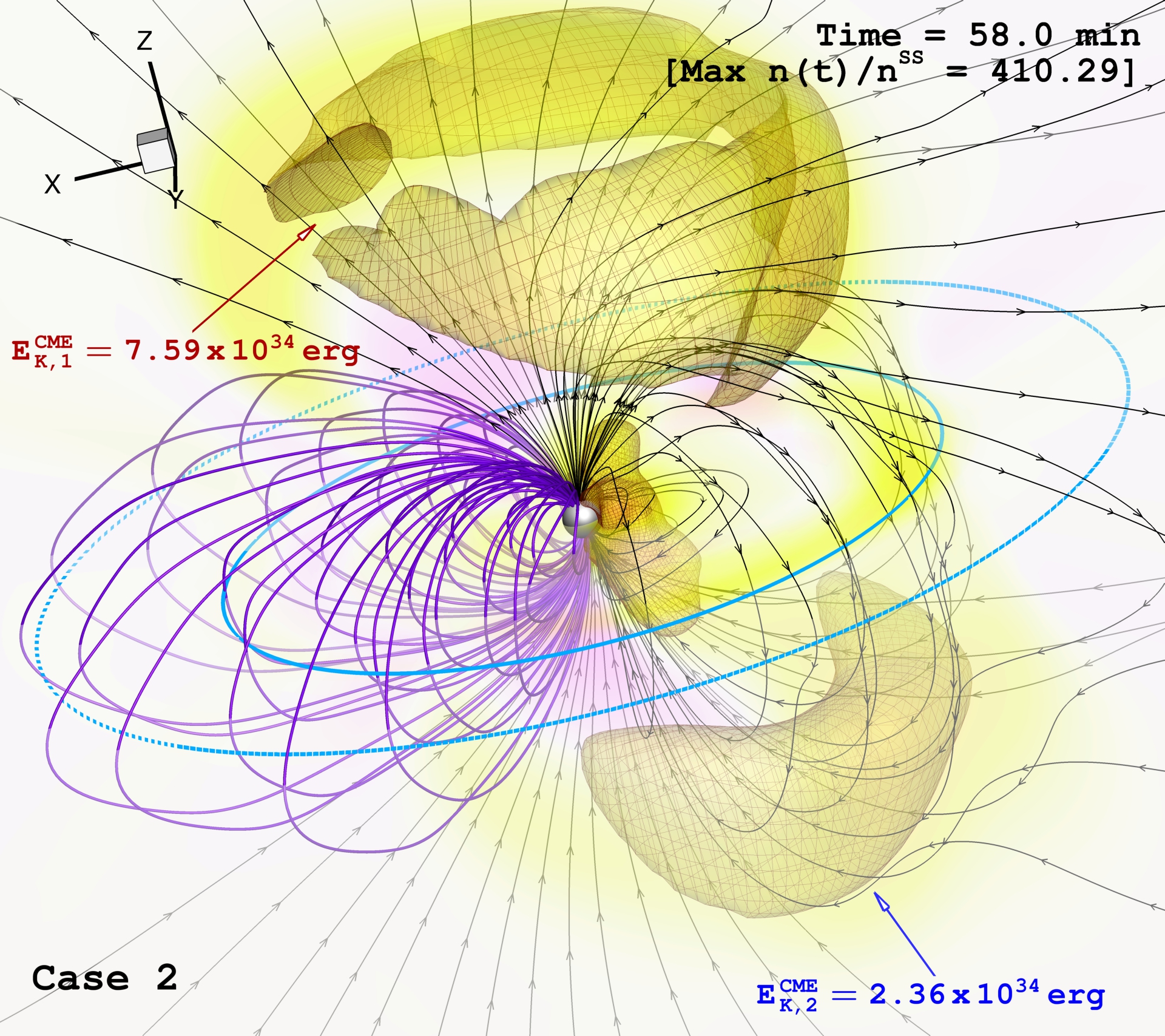}

\end{subfigure}\vspace{3pt}
 
\begin{subfigure}{0.495\linewidth}
\centering\includegraphics[trim = 0.1cm 0.1cm 0.1cm 0.1cm, clip=true,width=\textwidth]{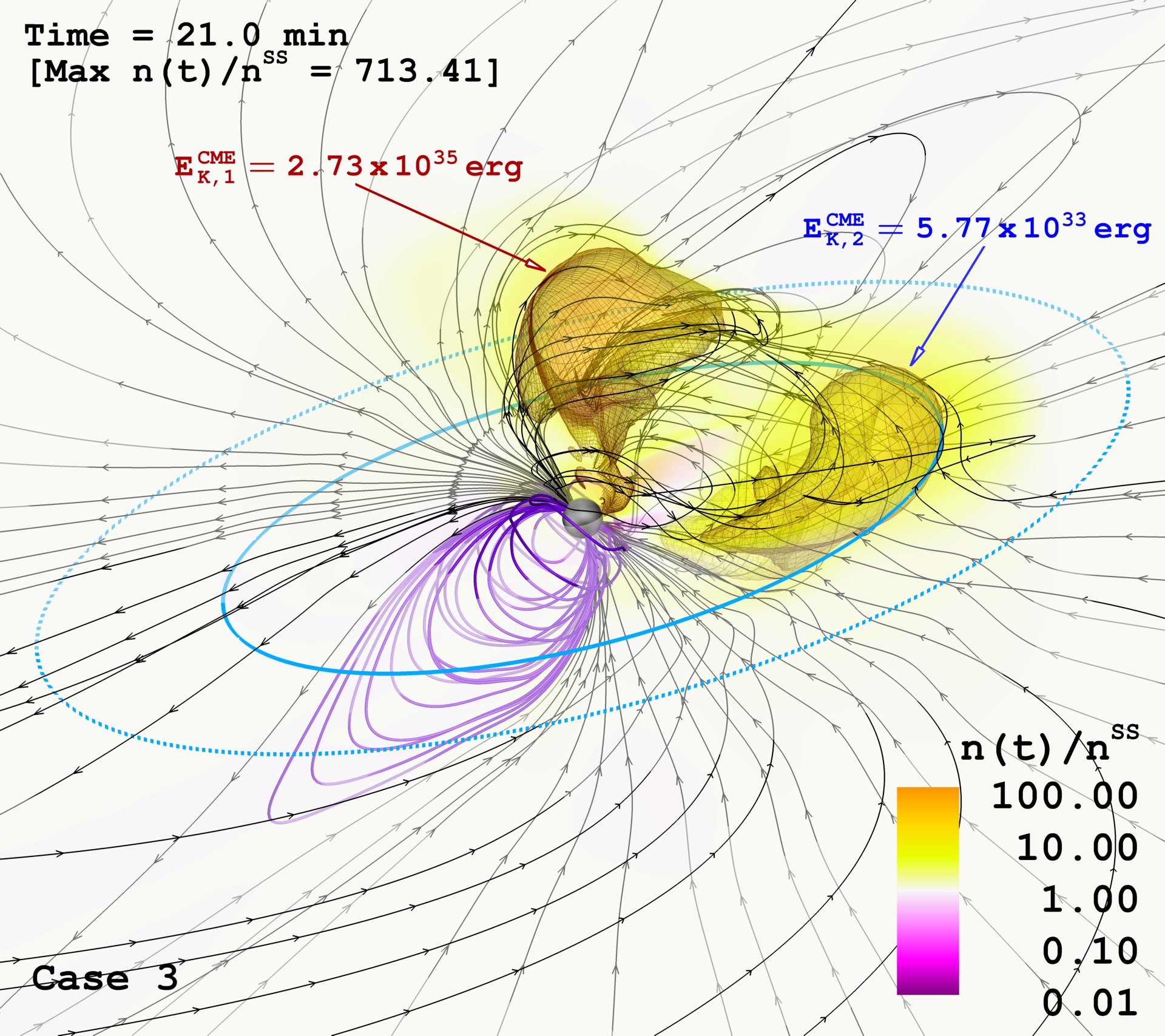}

\end{subfigure}%
\begin{subfigure}{0.495\linewidth}
\captionsetup{justification=centerlast,margin=0.25cm}
\captionof{figure}{\textbf{Figure 4:} Results of the CME simulations in the AU Mic system. The panels contain the emerging CMEs for each considered magnetic field/stellar wind case (Sects.~\ref{sec:Methods} and~\ref{sec:SS}). The color scale denotes the time-dependent density contrast $n(t)/n^{\rm SS}$, with the $n(t)/n^{\rm SS} = 10$ isosurface identifying the individual CME fragments. The snapshots show the perturbation arrival time at the orbit of AU~Mic~b (solid cyan line), with the instantaneous kinetic energy of each CME fraction indicated ($E_{\rm K, 1-2}^{\rm CME}$). Note that for Case 2 (top-right) the main fragments travel towards the polar regions and only a weaker perturbation arrives at the planets orbits. Selected magnetic field lines in the far (purple) and the erupting (black with arrows) sides are shown. The field of view is $60~R_{\bigstar}$. Animations of this figure are available.}
\end{subfigure}
\setcounter{figure}{4}
\caption{}\label{Fig_CME1}
\end{figure*}

\begin{figure*}
\begin{subfigure}{0.495\linewidth}
\centering\includegraphics[trim = 0.1cm 0.1cm 0.1cm 0.1cm, clip=true,width=\textwidth]{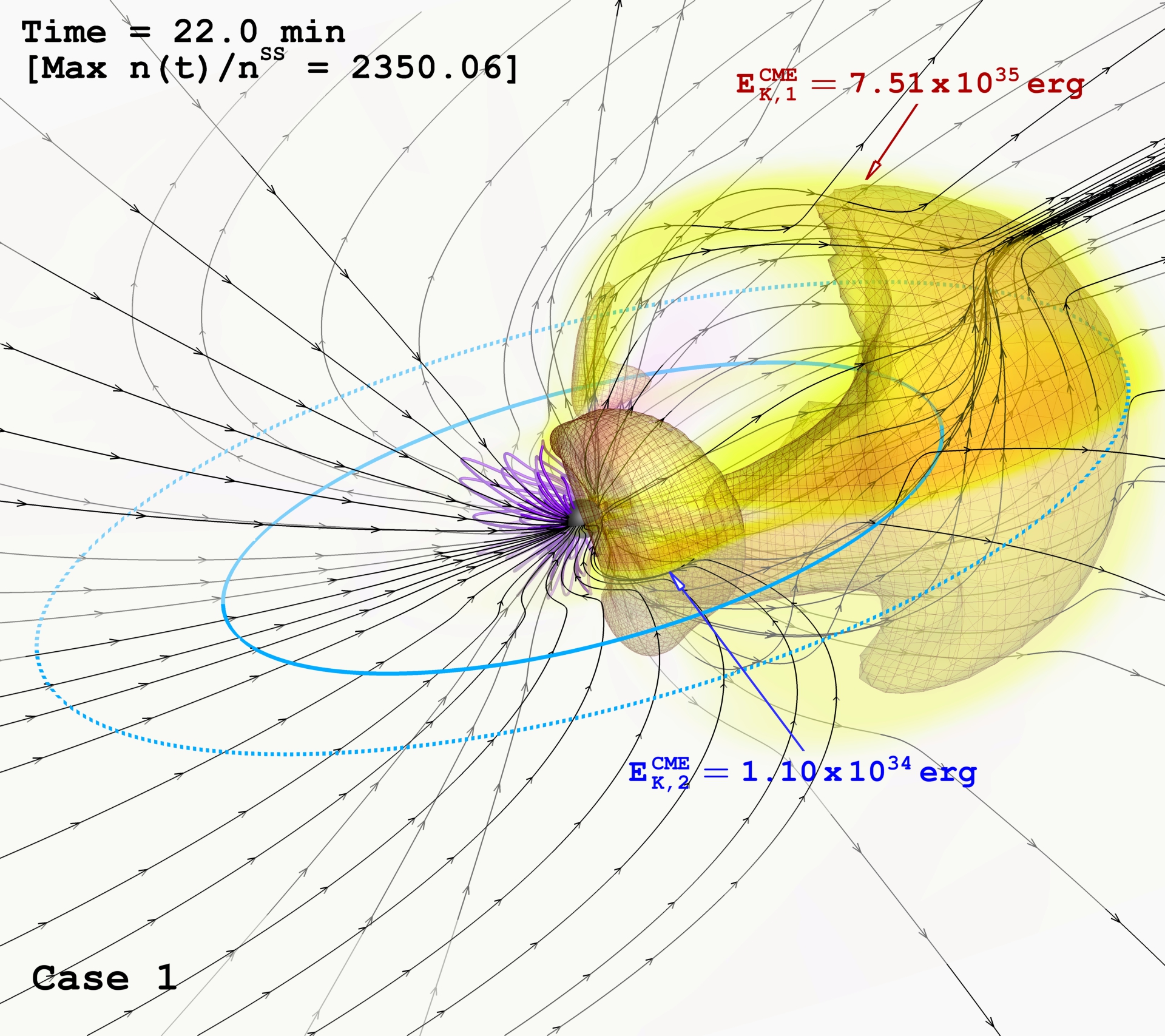}

\end{subfigure}
\begin{subfigure}{0.495\linewidth}
\centering\includegraphics[trim = 0.1cm 0.1cm 0.1cm 0.1cm, clip=true,width=\textwidth]{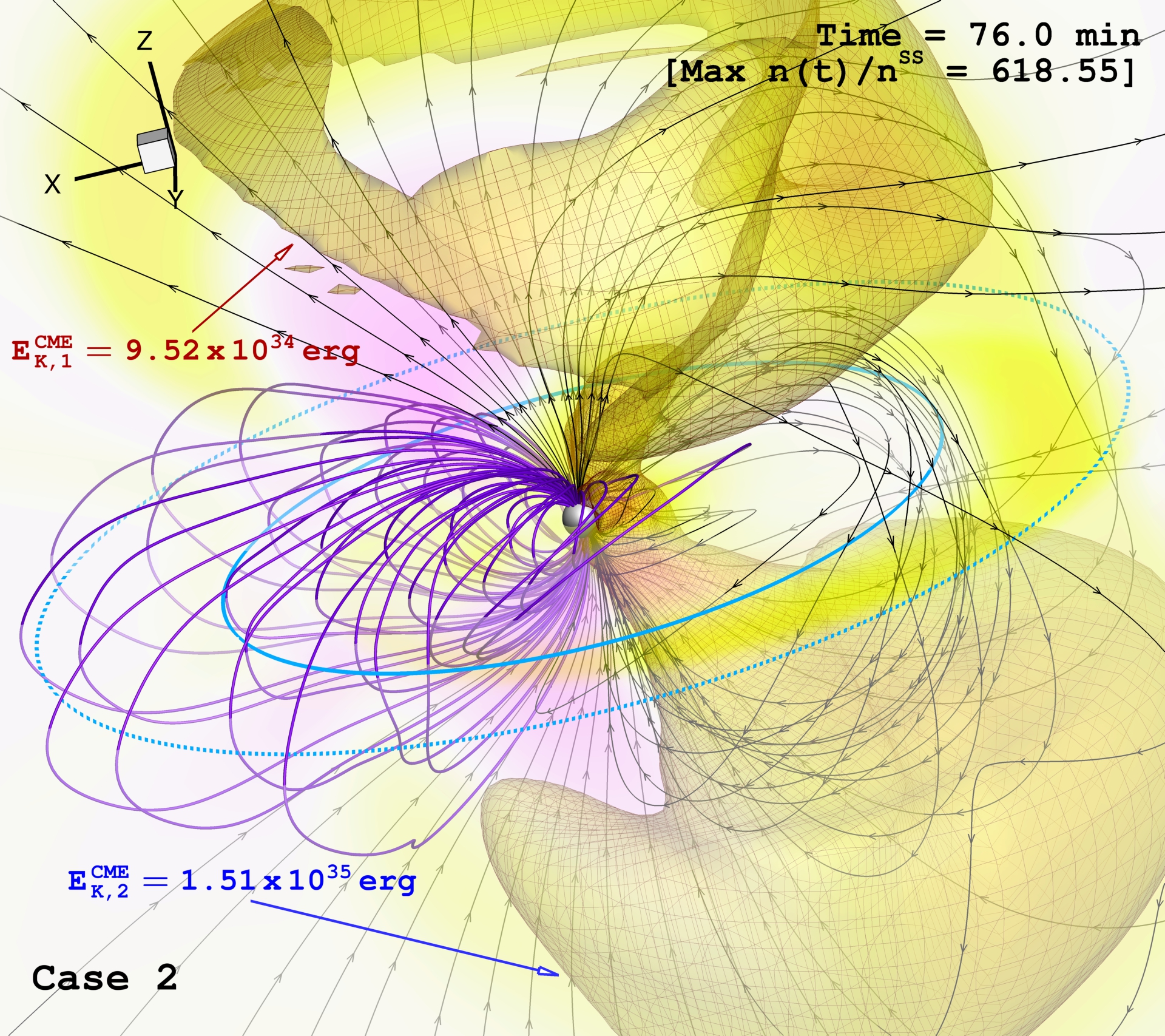}

\end{subfigure}\vspace{3pt}
 
\begin{subfigure}{0.495\linewidth}
\centering\includegraphics[trim = 0.1cm 0.1cm 0.1cm 0.1cm, clip=true,width=\textwidth]{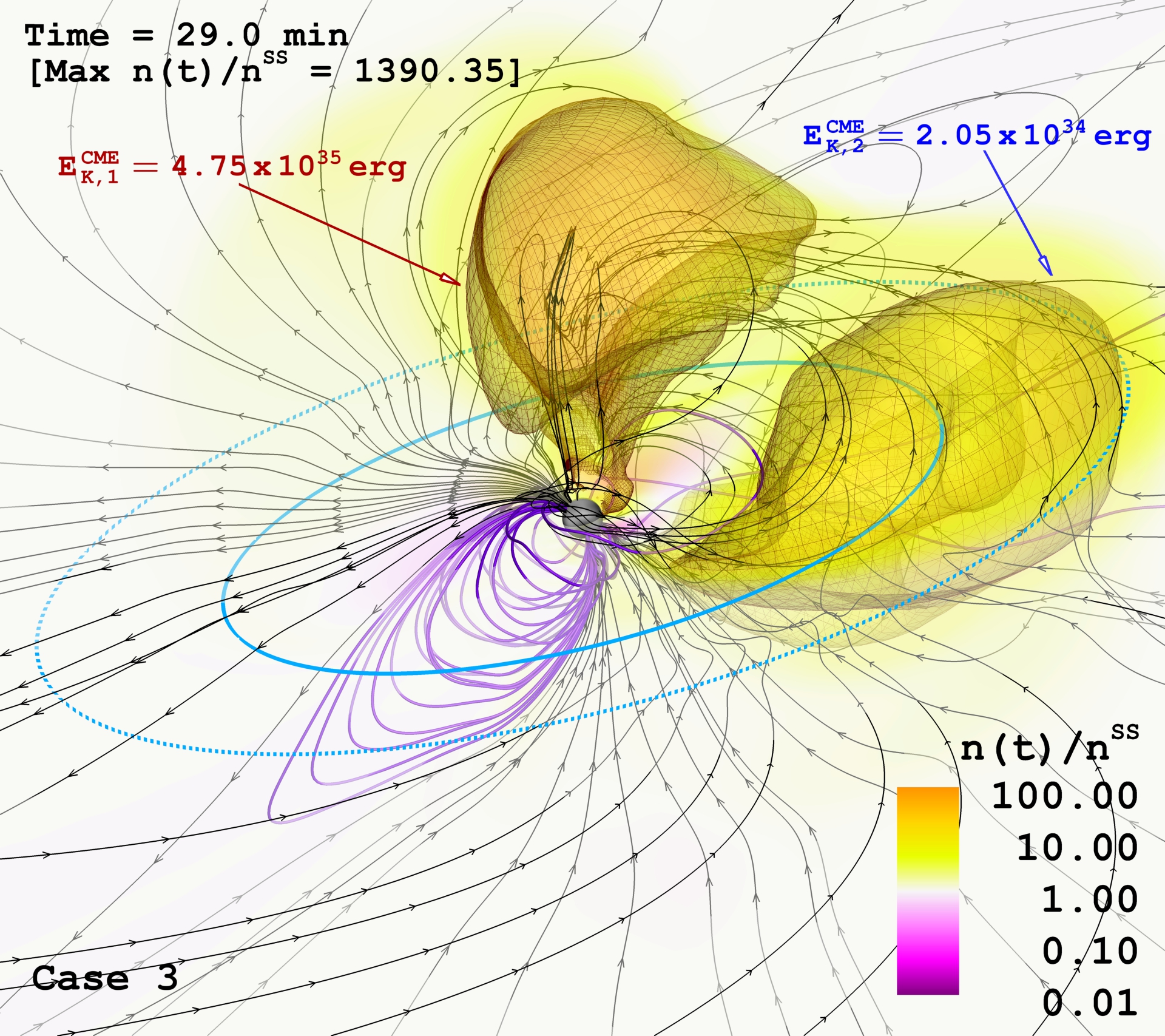}

\end{subfigure}%
\begin{subfigure}{0.495\linewidth}
\captionsetup{justification=centerlast,margin=0.25cm}
\captionof{figure}{\textbf{Figure 5:} Results of the CME simulations in the AU Mic system. See caption of Fig.~\ref{Fig_CME1}. The snapshots correspond to the perturbation arrival time at the orbit of AU~Mic~c (dashed cyan line), with the instantaneous kinetic energy of each CME fraction indicated ($E_{\rm K, 1-2}^{\rm CME}$). Animations of this figure are available.}
\end{subfigure}
\caption{}\label{Fig_CME2}
\end{figure*}

\subsection{Space weather forecast for AU Mic b and c}\label{sec:Planets}

\captionsetup[figure]{labelformat=default}
\begin{figure*}[!t]
\center 
\includegraphics[trim = 0.75cm 0.4cm 4.4cm 1.cm,clip=true,width=0.49\textwidth]{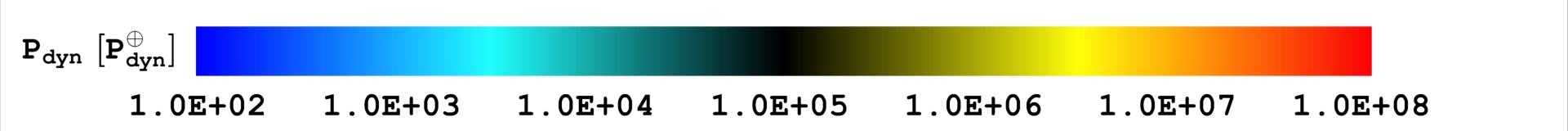} \includegraphics[trim = 0.75cm 0.4cm 4.4cm 1.cm,clip=true,width=0.49\textwidth]{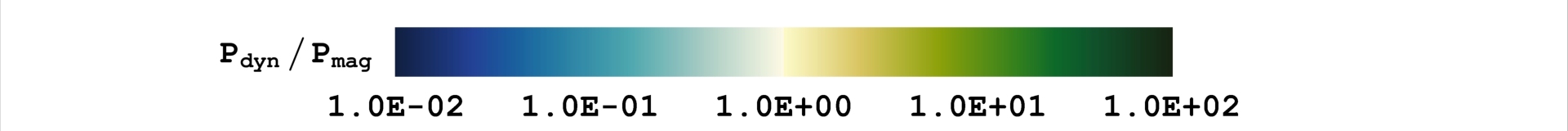}
\includegraphics[trim = 0.3cm 2.7cm 1.0cm 1.cm, clip=true,width=0.49\textwidth]{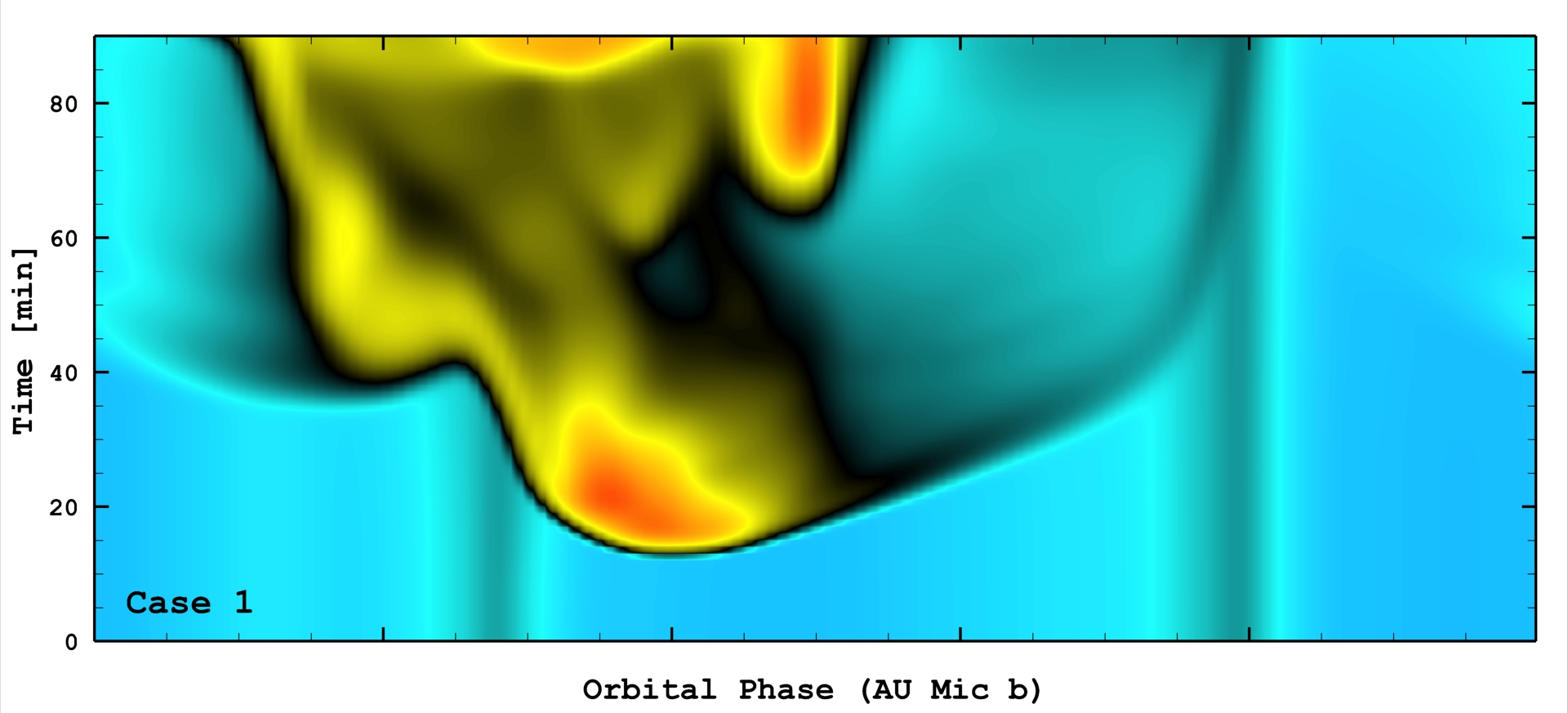} \includegraphics[trim = 0.3cm 2.7cm 1.0cm 1.cm, clip=true,width=0.49\textwidth]{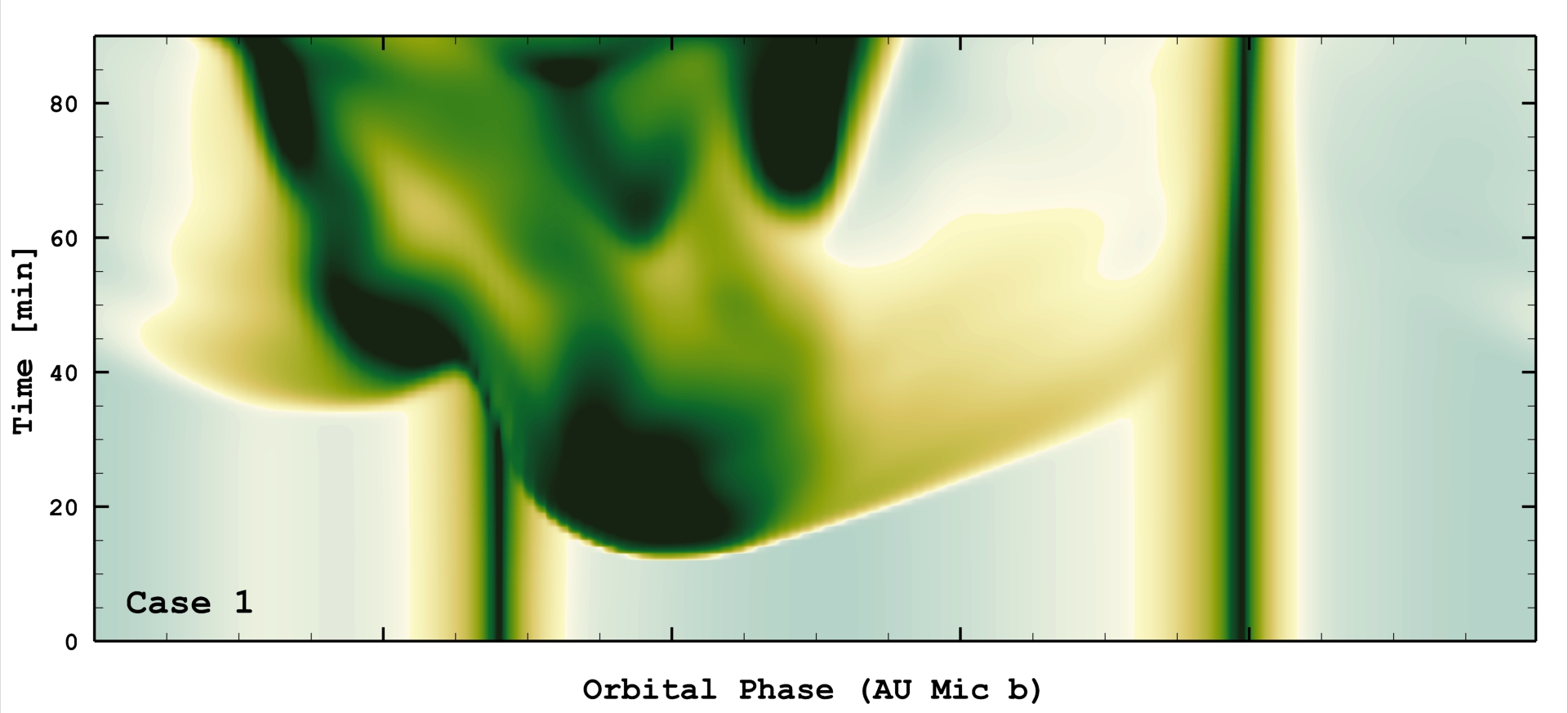}
\includegraphics[trim = 0.3cm 2.7cm 1.0cm 1.cm, clip=true,width=0.49\textwidth]{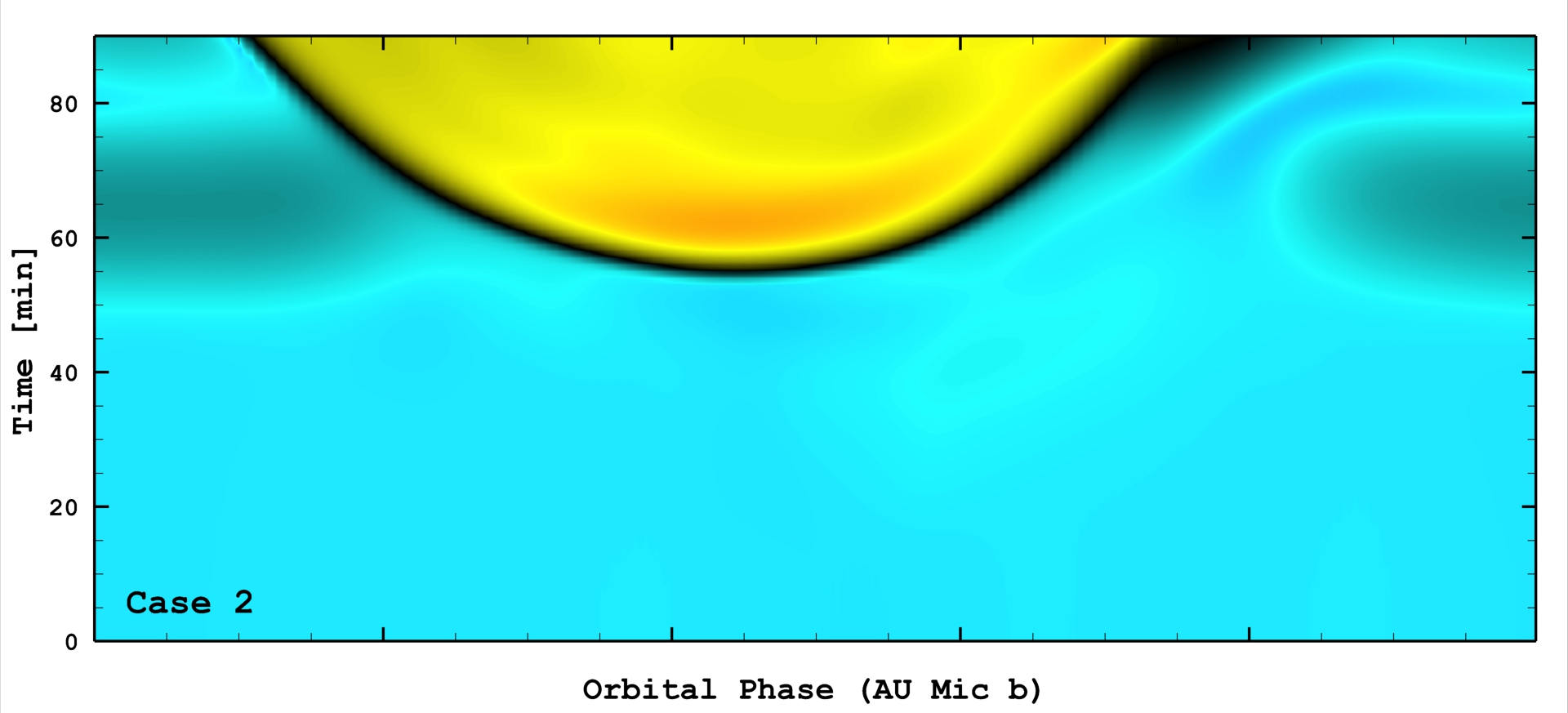} \includegraphics[trim = 0.3cm 2.7cm 1.0cm 1.cm, clip=true,width=0.49\textwidth]{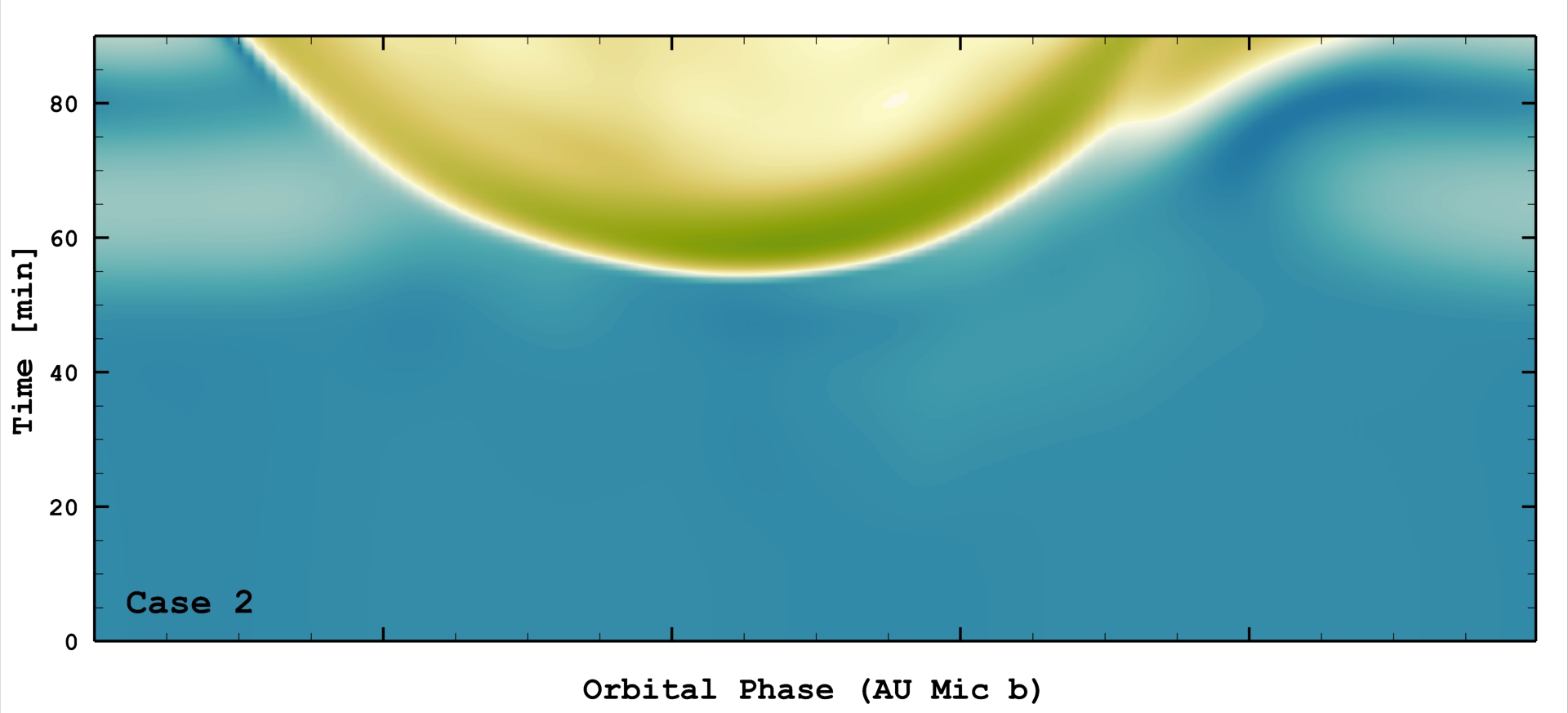}
\includegraphics[trim = 0.3cm 0.3cm 1.0cm 1.cm, clip=true,width=0.49\textwidth]{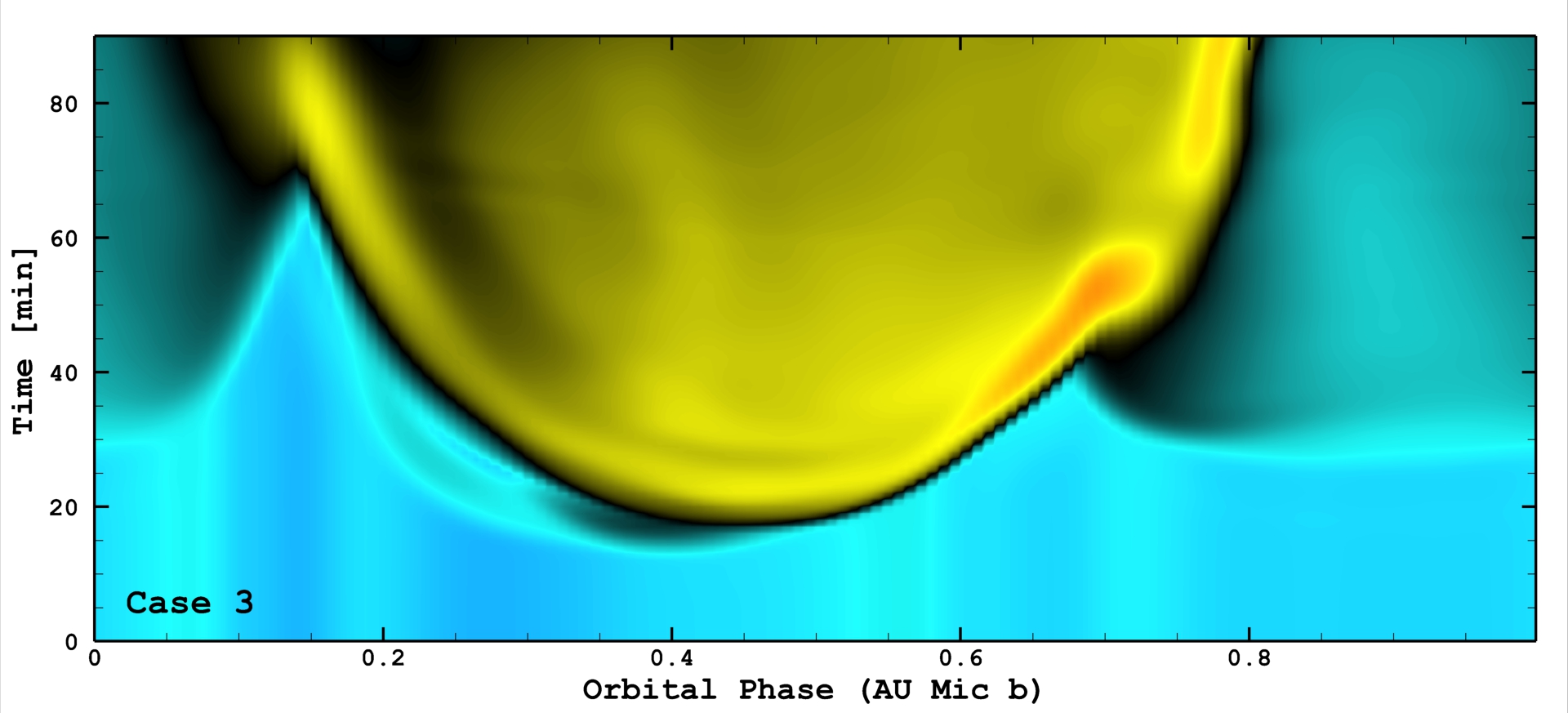} \includegraphics[trim = 0.3cm 0.3cm 1.0cm 1.cm, clip=true,width=0.49\textwidth]{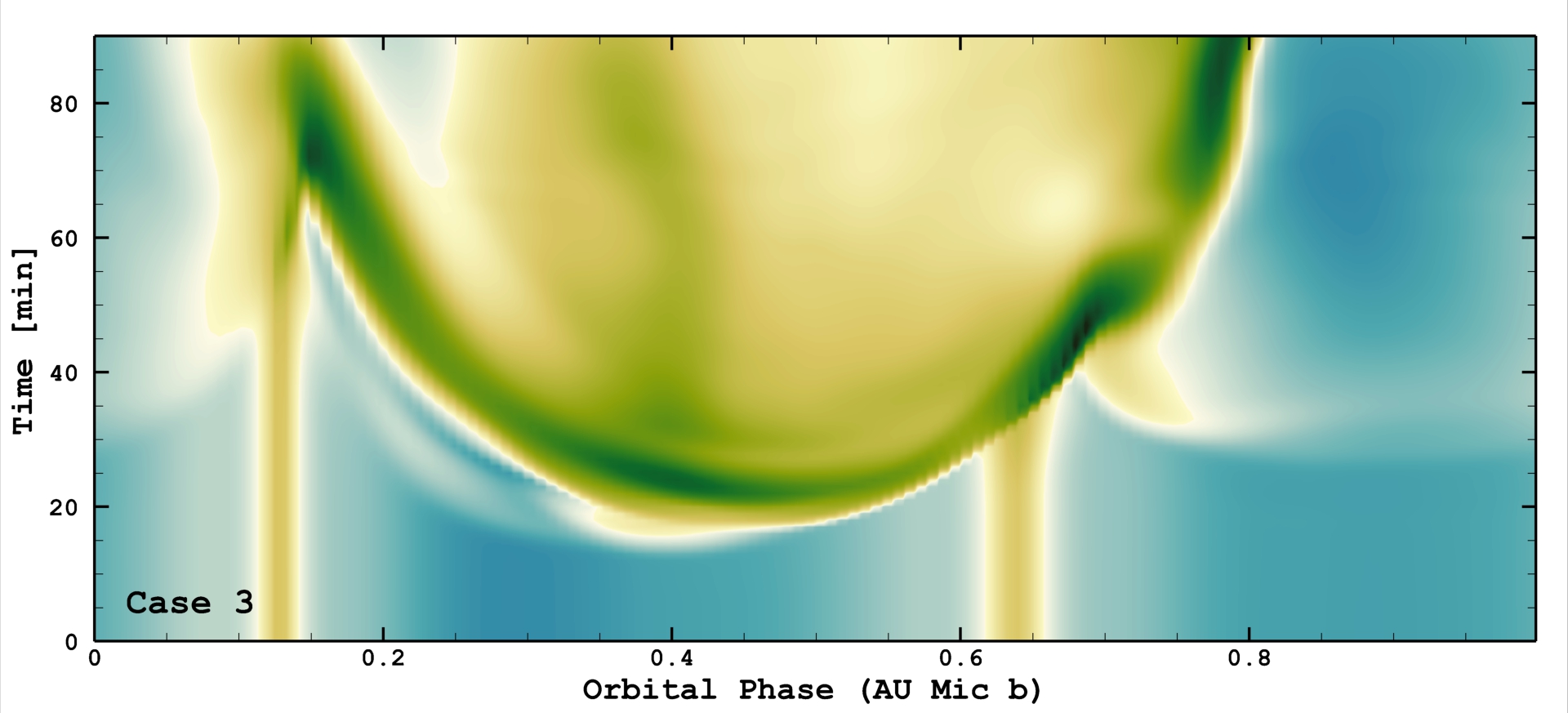}
\caption{Temporal evolution of the space weather conditions along the orbit of AU~Mic~b during the simulated CME events (Sect.~\ref{sec:CMEs}). The behaviour of $P_{\rm dyn}$ (left) and $P_{\rm dyn}/P_{\rm mag}$ (right) are shown for each considered case as indicated (see Fig.~\ref{Fig_CME1}). The values of $P_{\rm dyn}$ are normalized to the typical conditions experienced by the Earth ($P^{\oplus}_{\rm dyn} = 1.5$~nPa). The vertical features in Cases 1 (phases $0.25$ and $0.78$) and 3 (phases $0.10$ and $0.63$) correspond to the high-density stellar wind streamers associated with the current sheet.}
\label{SW_TD_AUMicb}
\end{figure*}

\begin{figure*}[!t]
\center 
\includegraphics[trim = 0.75cm 0.4cm 4.4cm 1.cm,clip=true,width=0.49\textwidth]{AUMic-b_Pdyn_ColorBar.jpeg} \includegraphics[trim = 0.75cm 0.4cm 4.4cm 1.cm,clip=true,width=0.49\textwidth]{AUMic-b_Prat_ColorBar.jpeg}
\includegraphics[trim = 0.3cm 2.7cm 1.0cm 1.cm, clip=true,width=0.49\textwidth]{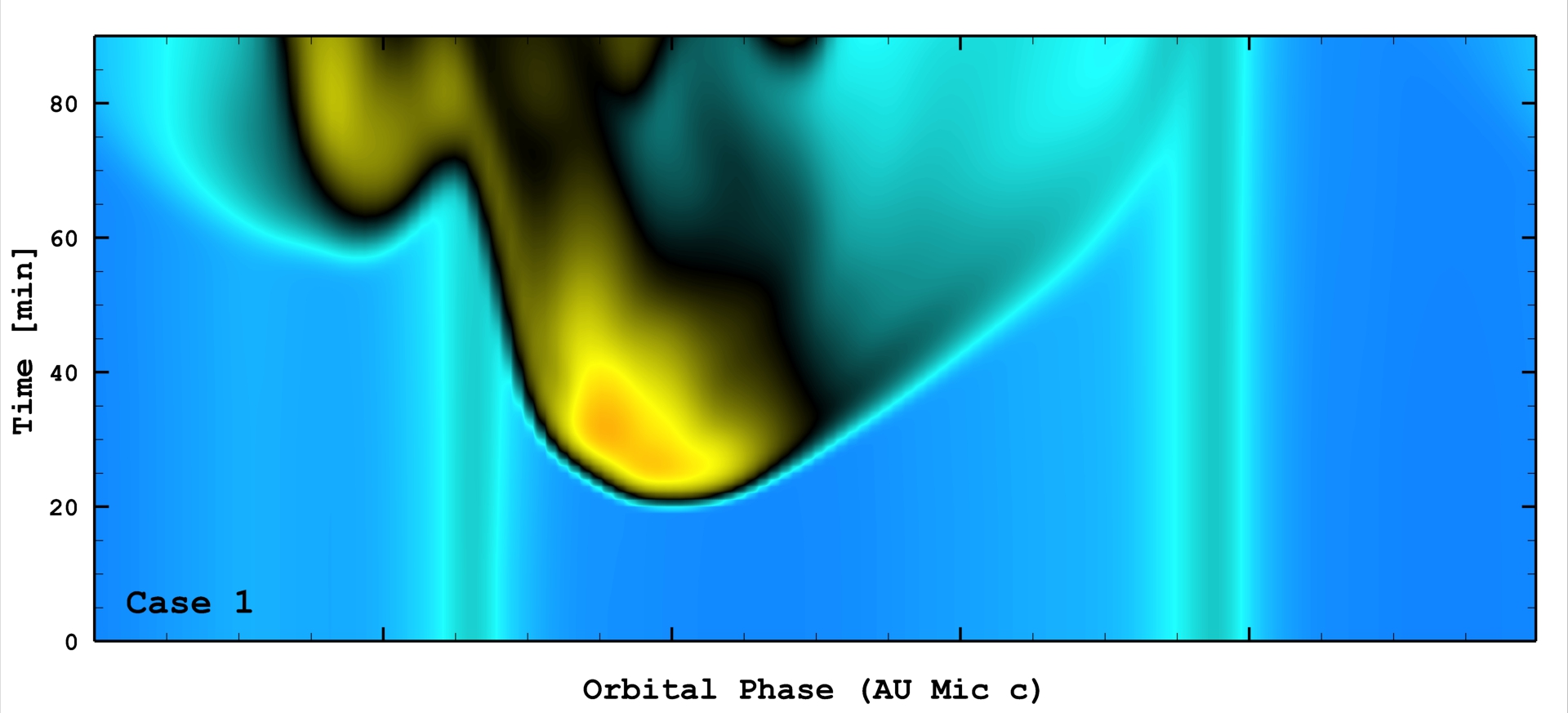} \includegraphics[trim = 0.3cm 2.7cm 1.0cm 1.cm, clip=true,width=0.49\textwidth]{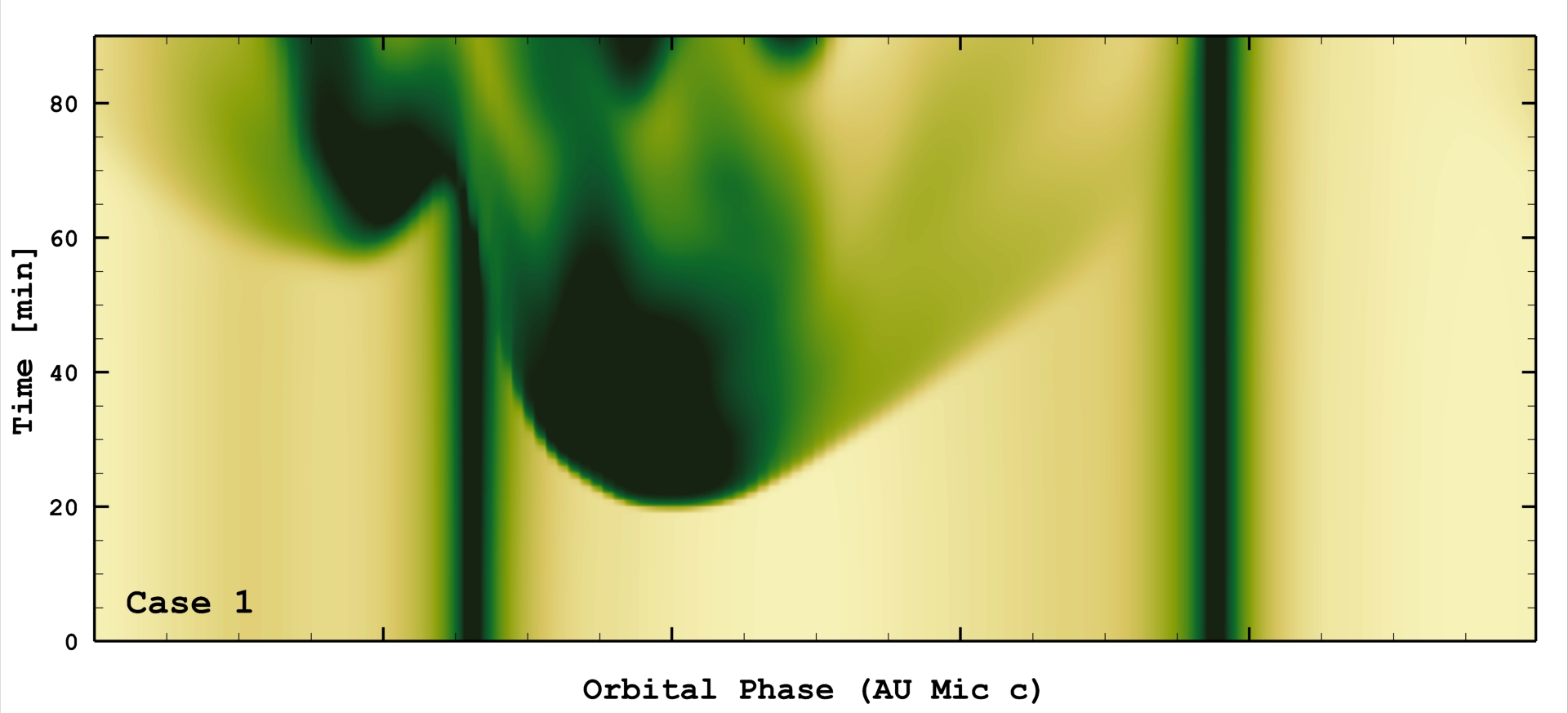}
\includegraphics[trim = 0.3cm 2.7cm 1.0cm 1.cm, clip=true,width=0.49\textwidth]{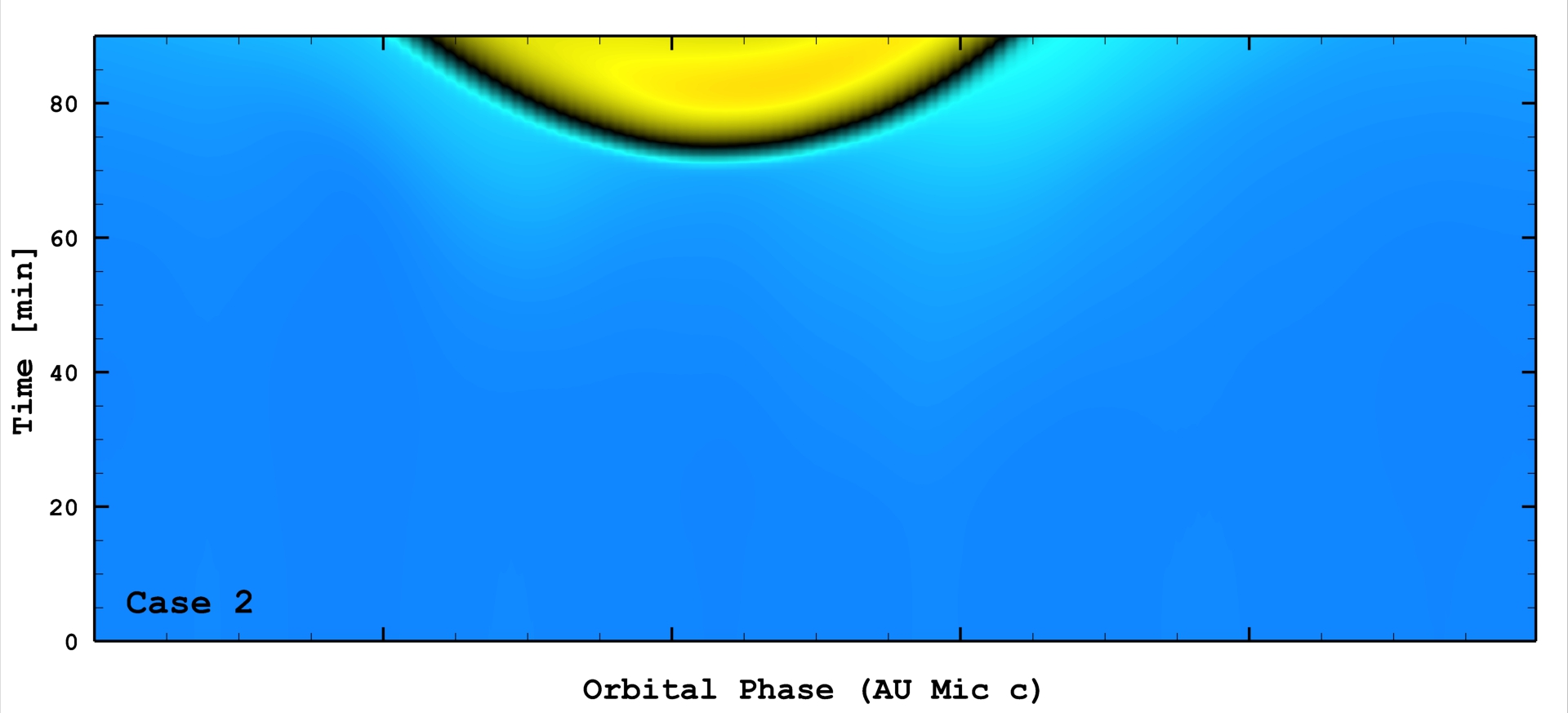} \includegraphics[trim = 0.3cm 2.7cm 1.0cm 1.cm, clip=true,width=0.49\textwidth]{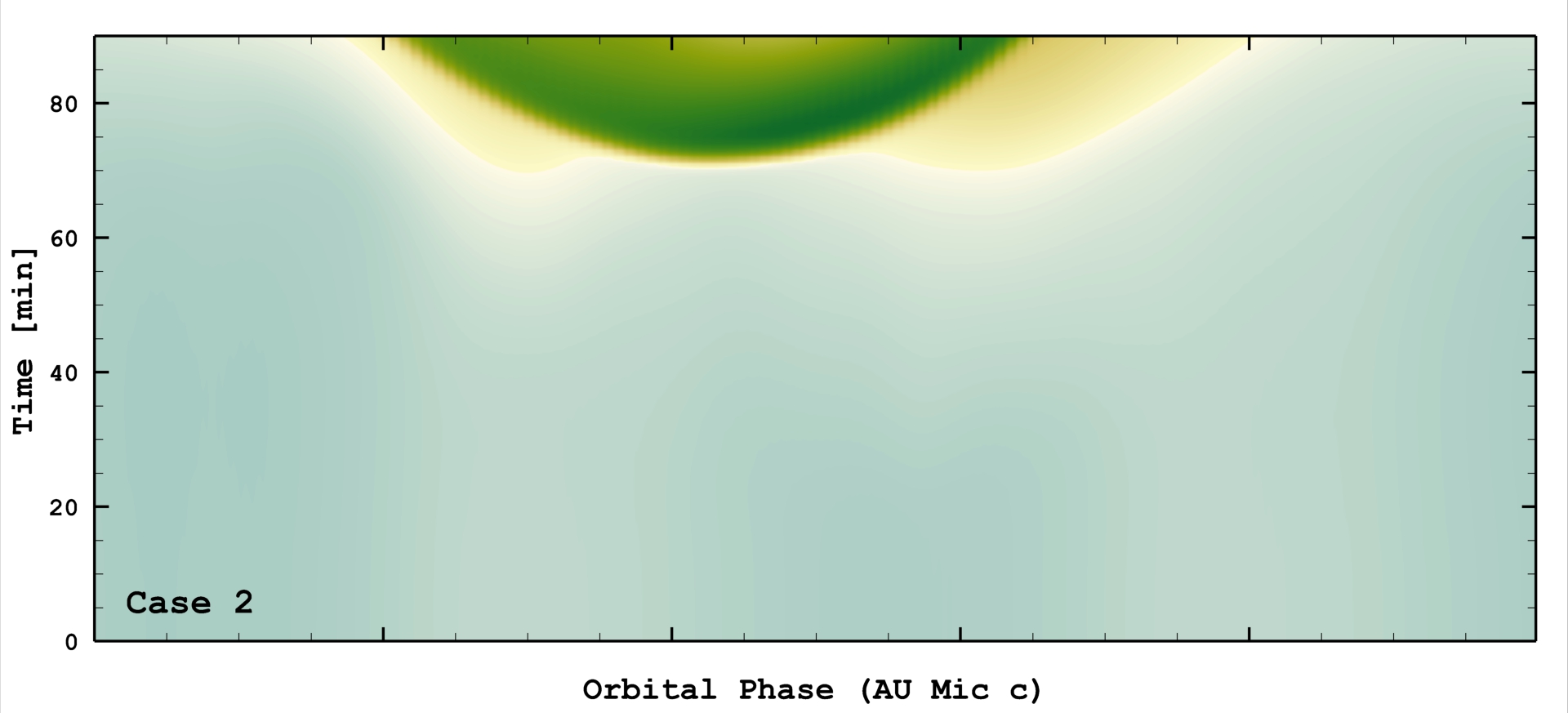}
\includegraphics[trim = 0.3cm 0.3cm 1.0cm 1.cm, clip=true,width=0.49\textwidth]{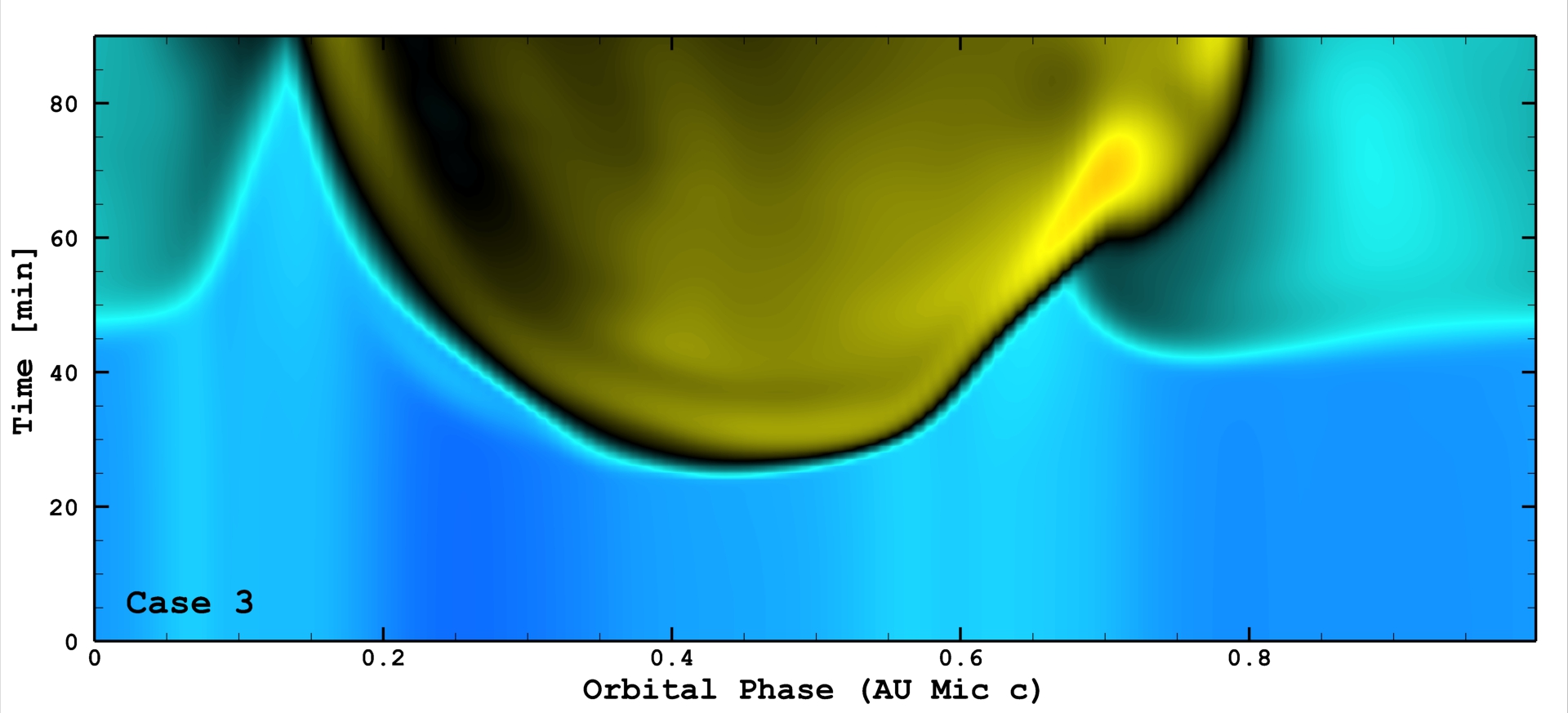} \includegraphics[trim = 0.3cm 0.3cm 1.0cm 1.cm, clip=true,width=0.49\textwidth]{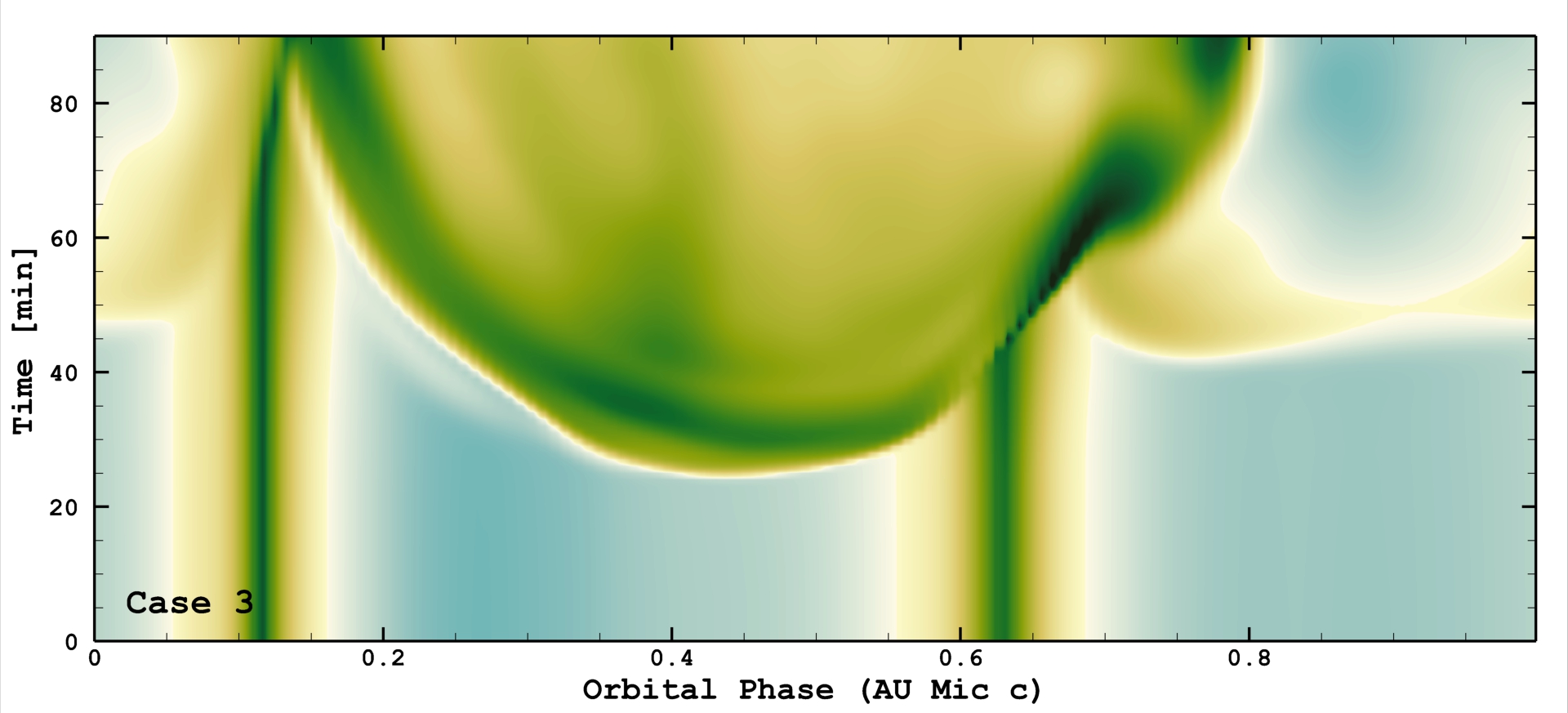}
\caption{Temporal evolution of the space weather conditions along the orbit of AU~Mic~c during the simulated CME events. See caption of Fig.~\ref{SW_TD_AUMicb}.}
\label{SW_TD_AUMicc}
\end{figure*}

\noindent Figure~\ref{SW_TD_AUMicb} shows the conditions along the orbit of AU~Mic~b  during the evolution of the exoplanet-effective CMEs emerging from each analyzed case. We include the behaviour of local dynamic pressure ($P_{\rm dyn}$, left) as well as its ratio to the magnetic pressure ($P_{\rm dyn}/P_{\rm mag}$, right). An analogous diagram along the orbit of AU~Mic~c is presented in Fig.~\ref{SW_TD_AUMicc}. The steady-state solutions (Fig.~\ref{Fig_SW}, Sect.~\ref{sec:SS}) provide the conditions at $t=0$~min in these time-phase representations.

Apart from differences in the CME speeds, the visualizations of Figs.~\ref{SW_TD_AUMicb} and \ref{SW_TD_AUMicc} reveal various noteworthy features. The CME arrival in all cases is characterized by an extremely sharp increase in $P_{\rm dyn}$ with respect to the pre-CME state\,---roughly of two to four orders of magnitude for planets b and c, respectively. While the CME conditions are harsher in absolute terms for the planet closer to the star, the milder steady-state stellar wind environment surrounding the outermost planet leads to a larger $P_{\rm dyn}$ jump at the CME arrival.

In the denser regions of the expanding CME front (forming the cores of the CME fragments reaching the planets\footnote[9]{In the Case~1 simulation, the arrival of the second CME fragment at the orbit of AU~Mic~b can be seen around $t\simeq70$~min (Fig~\ref{SW_TD_AUMicb}, top-left panel).}; see Figs.~\ref{Fig_CME1} and \ref{Fig_CME2}), $P_{\rm dyn}$ rises to values up to $10^{8}$ times the dynamic pressure experienced by the Earth on nominal solar wind conditions. Even the arrival of a Carrington-type CME event at the Earth's orbit would induce a $P_{\rm dyn}$ transient five orders of magnitude smaller\footnote[10]{Assuming a density contrast of $n(t)/n^{\rm SS} = 20$ and a CME speed of $3000$~km~s$^{-1}$ (see \citeads{2013SpWea..11..585B}).}. While such $P_{\rm dyn}$ values also exceed by far the total pressure ($P_{\rm dyn} + P_{\rm mag}$) predicted for planets orbiting low-mass stars at much closer distances compared to the AU~Mic planets (e.g.,~\citeads{2015MNRAS.449.4117V}, \citeads{2017ApJ...843L..33G}), these extreme CME-driven conditions persist only for a relatively short time (tens of minutes for the events considered here). Therefore, their global impact on a given atmospheric and magnetospheric structure of an exoplanet will mainly depend on the effective escaping CME rate of the star (mediated by the stellar large-scale magnetic field).

The bulk of the eruption strongly affects a large fraction of the orbit of both exoplanets\,---between $60$\% to $100$\% depending on the case. In those regions the  dynamic pressure ranges from $10^{6}~P^{\oplus}_{\rm dyn}$ down to a few $10^{4}~P^{\oplus}_{\rm dyn}$. As illustrated in the right panels of Figs.~\ref{SW_TD_AUMicb} and \ref{SW_TD_AUMicc}, these values are sufficient to overcome the general dominance of $P_{\rm mag}$ in this system. Furthermore, the CME-driven enhancement in the local plasma density and velocity will shift the conditions along the affected orbital phases from the sub-Alfv\'enic to the super-Alfv\'enic regime (see Sect.~\ref{sec:SS}). However, this situation is expected to be temporary, as the sub-Alfv\'enic conditions along the orbit return once the stellar wind has relaxed into a new steady-state (not captured in our simulations).

One interesting and counter-intuitive consequence of this CME-driven transition is that the eruption passage would shutdown any possible planet-induced stellar radio emission via the electron cyclotron maser instability (ECMI, see~\citeads{2013A&A...552A.119S}, \citeads{2021MNRAS.504.1511K}). This type of emission is expected to occur in star-planet interactions (SPI) within the sub-Alfv\'enic regime of the stellar wind, where Alfv\'en waves traveling towards the star would carry enough energy to induce radio emission via ECMI in the stellar corona \citepads{2018ApJ...854...72T}. As the sub-Alfv\'enic regime is temporarily lifted by the CME passage, the dimming of this radio signal (assuming that a pre-CME baseline could be detected and isolated from other possible stellar sources) may potentially serve to find candidate CME events on the host star\,---analogous to coronal dimming events observed in the Sun and other stars (see \citeads{2016ApJ...830...20M}, \citeads{2021NatAs...5..697V}). Furthermore, if such a CME event would be associated with a flare, the radio dimming time-lag should be able to yield an estimate of the CME speed (provided that the orbit of the exoplanet is known). While their characterization is beyond the scope of this paper, these SPI-ECMI radio dimming features could constitute an alternative pathway to study the eruptive behaviour of exoplanet hosts. This is not only relevant as the information on stellar CMEs in general is very scarce (see~\citeads{2019ApJ...877..105M}, \citeads{2019NatAs...3..742A}), but also since the radio properties of cool stars at low frequencies are starting to be unveiled using ground-based observations (\citeads{2020NatAs...4..577V}, \citeads{2021NatAs.tmp..196C}).    

Finally, as noted in Sect.~\ref{sec:intro}, observations have shown considerable flare activity in AU~Mic, with estimated rates in the range of $5.54 - 6.35$~flares/d for events with bolometric energies $E^{\rm FL} < 10^{32}$~erg, down to one event every 10 days for large flares with $E^{\rm FL} \sim 10^{34}$~erg (\citeads{2021A&A...649A.177M}, \citeads{2021arXiv210903924G}). As such, on a given rotation AU~Mic displays on the order of $25-30$ flaring events with $E^{\rm FL} < 10^{32}$~erg, and about 5 flares that reach $10^{33}$~ergs in bolometric energy.  Even though CMEs associated with this flare activity will suffer strong magnetic suppression, a quiescent state in AU~Mic (described in Sect.~\ref{sec:SC} and also by other authors in previous studies), might not represent the most common conditions around this flare star and its exoplanets. Such conditions are expected to be characterised by a rapidly changing corona and stellar wind environment, with considerable variability on a sub-rotational period time-scale. 

While the flare-CME candidate inspiring our simulations will not be as frequent as those more milder events\footnote[11]{Extrapolating the flare frequency distribution from \citetads{2021arXiv210903924G} to the estimated energy of the event ($E^{\rm FL} > 10^{35}$~erg, \citeads{1999ApJ...510..986K}), yields a cumulative rate $<1$~flare/100~d.}, its evolution reveals how much the space weather of the exoplanets and the entire inner astrosphere can be affected by these energetic transients. The resulting CME fragmentation leads to a wide range of latitudes, including the polar regions, being disturbed by the expanding structures (Figs.~\ref{Fig_CME1} and \ref{Fig_CME2}). Changes in the geometry of the AS as well as the complete disruption of the current sheet (see Figs.~\ref{SW_TD_AUMicb} and~\ref{SW_TD_AUMicc}), are just two examples of the CME influence on the circumstellar environment. As the typical time-scale involved in all of these CME-related affectations is comparable to the exoplanets' transit times ($3.56$~h for AU~Mic~b and $4.42$~h AU~Mic~c, \citeads{2021arXiv210903924G}), any observational and/or modelling efforts performed on this object should include time-dependent effects in their analysis. We will pursue this research direction in two complementary numerical studies on AU~Mic, evaluating the influence of this extreme CME on atmospheric outflow patterns from the innermost planet (\textcolor{blue!60!black}{Cohen et al. in prep.}), as well as investigating the pre- and post-CME energetic particle environments in this system (\textcolor{blue!60!black}{Fraschetti et al. in prep.}).  

\section{Summary and Conclusions}\label{sec:Conclusions}

\noindent In this study, we employed state-of-the-art numerical models to simulate the space weather of the planet-hosting flare star AU~Mic. Two coupled models are used to characterize the quiescent stellar wind and the temporal evolution of a very energetic CME event in this system. Three different configurations of the large-scale magnetic field of AU~Mic (inspired by Zeeman-Doppler Imaging observations), as well as eruption parameters sufficient to power the best CME candidate in this star, were incorporated as boundary conditions in our simulations. 

The analyses of the steady-state and transient stellar wind conditions around the star were performed separately. Using the former, we showed that our models predict mass loss rates in line with previous theoretical and numerical model expectations for AU~Mic, and with current observational constraints for other M-dwarfs including a measurement available for the flare star EV~Lac (relatively similar to AU~Mic in spectral type, rotation period and activity levels). 

Our results indicate that the planets of the AU~Mic system experience extreme space weather conditions. In agreement with previous studies, we found that both exoplanets lie inside the sub-Alfv\'enic region of the stellar wind for the large majority of their orbits. At their respective orbital separation, AU~Mic~b and AU~Mic~c endure stellar wind dynamic pressures between two and four orders of magnitude larger than the average value experienced by the Earth due to the solar wind. However, the conditions surrounding the AU~Mic planets are most likely dominated by the local magnetic pressure, which can surpass the dynamic pressure of the stellar wind by relatively large factors (around $5 - 10$ for some of the cases considered here).     

For our modelling of the extreme CME event, a time-dependent simulation (initialized from each of the three ZDI-driven steady-state solutions) followed the evolution of an erupting twisted magnetic flux-rope, which was anchored at the inner boundary of the domain. We found that, despite having identical parameters, the flux-rope eruption generated different CMEs in terms of geometry, mass, velocity, and kinetic energy, for the different starting wind solutions. These differences are attributed to the properties of large-scale magnetic field (strength and topology) which also dictate the structure of the stellar wind in which the CMEs develop. In particular, the varying levels of magnetic suppression resulted in CMEs with global radial speeds between $5000 - 10000$~km s$^{-1}$, masses close to $2\times10^{18}$~g, and kinetic energies within the $10^{35} - 10^{36}$~erg range.    

One common feature among all the simulated events was the fragmentation of the CME due to the confining influence of the large-scale field. Two main fragments emerged in each of our CME simulations, whose properties were isolated and compared with the behaviour of the CME events as a whole. This analysis revealed that the re-distribution of the total CME mass and kinetic energy among the fragments is not uniform, resulting in differences of more than one order of magnitude for these parameters. Additionally, while the launching location of the flux-rope was near the stellar equator, the fragmentation of the eruption led to pole-ward CMEs in two out of the three cases considered. While further investigation is required, this situation might be connected with the presence of a ``guiding'' anti-symmetric component in the large-scale magnetic field (absent in the case lacking these high-latitude CMEs).     

Our simulations showed that the CME fragmentation also plays a major role in determining the impact of the eruption on the AU~Mic exoplanets. In the two cases displaying pole-ward CMEs, the perturbations arriving at the exoplanet orbits (aligned with the equatorial plane) were significantly less energetic compared with their high-latitude counterparts (differences in the CME kinetic energy of up to two orders of magnitude). From a purely CME dynamics point of view, if an anti-symmetric large-scale magnetic field indeed enables a pole-ward fragmentation of the eruption, such magnetic topology could be beneficial for close-in exoplanets orbiting active stars. One particular case of such anti-symmetric field would be a large-scale dipole aligned with the stellar rotation axis. This configuration would have the additional benefit of a smaller stellar wind Alfv\'en surface at the equatorial plane, allowing super-Alfv\'enic orbits closer to the star with the possibility of closed magnetospheres around the planets (not possible for sub-Alfv\'enic orbits).

Finally, we also analyzed the effects along the exoplanet orbits due to the arrival of the exoplanet-effective CME fractions (traveling close to the equatorial plane). In all the considered cases, these eruptions completely altered the space weather conditions for large portions of the exoplanet orbit (between 60\% to 100\%). Their impact yielded dynamical pressure increases of between four to six orders of magnitude with respect to the steady-state, with the largest increase being associated with the high-density CME fragment cores. The CME-induced dynamic pressure transient surpassed the local magnetic tension by large factors (between $10$ and $>10^{2}$), temporarily shifting the exoplanetary conditions from sub-Alfv\'enic to super-Alfv\'enic. We briefly explored one possible implication of this transition, where the CME arrival at the exoplanet orbit would induce a radio dimming event in the electron cyclotron maser instability emission from star-planet interactions. This feature, observable in principle using ground-based instrumentation, could open a window for future exploration of the still elusive CME behaviour in other stars. 



\acknowledgments
\noindent JJD and CG were funded by NASA contract NAS8-03060 to the CXC and thank the Director, Patrick Slane, for continuing advice and support. OC was supported by NASA NExSS grant NNX15AE05G. FF was supported, in part, by NASA under Grants NNX16AC11G and Chandra Theory Award Number $TM0-21001X$, $TM6-17001A$ issued by the Chandra X-ray Observatory Center, which is operated by the Smithsonian Astrophysical Observatory for and on behalf of NASA under contract NAS8-03060. KP and JC acknow\-led\-ge funding from the German \textit{Leibniz Gemeinschaft} under project number P67-2018. LMH has received funding from the European Research Council (ERC) under the European Union’s Horizon 2020 research and
innovation programme (grant agreement No. 853022, PEVAP). This work used SWMF/BATSRUS tools developed at The University of Michigan Center for Space Environment Modeling. 

%



\software{SWMF \citepads{2018LRSP...15....4G}}

\bibliographystyle{aasjournal}
\bibliography{Biblio}



\end{document}